\documentclass[10pt,conference]{IEEEtran}

\usepackage{caption}

\makeatother

\usepackage{url}
\usepackage{multirow}
\usepackage{color}
\usepackage{amsmath}
\usepackage{amssymb}
\usepackage{graphicx}
\usepackage{adjustbox}  
\usepackage{subcaption} 
\usepackage{color}
\usepackage{multirow}
\usepackage{algorithm}
\usepackage{algorithmic}
\usepackage{soul}
\usepackage{todonotes}
\usepackage{flushend}
\usepackage{comment}

\usepackage{float}
\let\labelindent\relax
\usepackage{enumitem}
\usepackage[normalem]{ulem} 
\useunder{\uline}{\ul}{} 

\usepackage{dblfloatfix}

\usepackage[square,sort,comma,numbers]{natbib}
\usepackage{caption}
\usepackage{xcolor,colortbl}

\usepackage{tikz}

\begin{document}

\title{{The AI Companion in Education: Analyzing the Pedagogical Potential of ChatGPT in Computer Science and Engineering}}






\title{{\huge The AI Companion in Education: Analyzing the Pedagogical Potential of ChatGPT in Computer Science and Engineering}}

\author{\IEEEauthorblockN{Zhangying He}
\IEEEauthorblockA{
\textit{California State University, Long Beach} \\ 
Long Beach, CA, USA \\}
\and

\space\space
\and

\space\space
\and

\IEEEauthorblockN{Thomas Nguyen}
\IEEEauthorblockA{
\textit{California State University, Long Beach}\\
 Long Beach, CA, USA\\}
\and

\space\space
\and

\space\space
\and

\IEEEauthorblockN{Tahereh Miari}
\IEEEauthorblockA{
\textit{Claremont Graduate University} \\ 
Claremont, CA, USA\\}
\and

\space\space
\and

\space\space
\and

\IEEEauthorblockN{Mehrdad Aliasgari}
\IEEEauthorblockA{
\textit{California State University, Long Beach} \\
 Long Beach, CA, USA \\}
\and

\space\space
\and

\space\space
\and

\space\space
\and

\space\space
\and

\IEEEauthorblockN{Setareh Rafatirad}
\IEEEauthorblockA{
\textit{University of California, Davis} \\
Davis, CA, USA \\}
\and
\space\space
\and

\space\space
\and

\space\space
\and

\space\space
\and

\IEEEauthorblockN{Hossein Sayadi}
\IEEEauthorblockA{
\textit{California State University, Long Beach} \\ 
Long Beach, CA, USA \\ \vspace{3ex}}
}

\maketitle

\footnotetext[1]{ This is the authors’ version of the work. The definitive version of record will be published in the \textit{2024 IEEE Global Engineering Education Conference (EDUCON'24)} proceedings.}

\begin{abstract}

Artificial Intelligence (AI), 
with ChatGPT as a prominent example, has recently taken center stage in various domains including higher education, particularly in Computer Science and Engineering (CSE). The AI revolution brings both convenience and controversy, offering substantial benefits while lacking formal guidance on their application. 
The primary objective of this work is to comprehensively analyze the pedagogical
potential of ChatGPT in CSE education, understanding its strengths and limitations from the perspectives of educators and learners. We employ a systematic approach, creating a diverse range of educational practice problems within CSE field, focusing on various subjects such as data science, programming, AI, machine learning,  networks, and more. 
According to our examinations, certain question types, like conceptual knowledge queries, typically do not pose significant challenges to ChatGPT, and thus, are excluded from our analysis. Alternatively, we focus our efforts on developing more in-depth and personalized questions and project-based tasks. 
These questions are presented to ChatGPT, followed by interactions to assess its effectiveness in delivering complete and meaningful responses. 
To this end, we propose a comprehensive five-factor reliability analysis framework to evaluate the responses.
This assessment aims to identify when ChatGPT excels and when it faces challenges. Our study concludes with a correlation analysis, delving into the relationships among subjects, task types, and limiting factors. This analysis offers valuable insights to enhance ChatGPT’s utility in CSE education, providing guidance to educators and students regarding its reliability and efficacy.

\end{abstract}

\begin{IEEEkeywords}
ChatGPT, Computer Science and Engineering, Education, Generative Artificial Intelligence, Reliability Analysis. 
\end{IEEEkeywords}

\section{Introduction}
\label{sec:intro}
Recent advancements in the field of Artificial Intelligence (AI) have shown significant promise across various applications,  notably in augmenting the quality of education \cite{Chung-EDUCON2022, FIE-2023}. 
In recent decades, Natural Language Processing (NLP), as a pivotal domain within AI, 
has experienced remarkable growth. This evolution is marked by significant strides in neural network architectures and learning methodologies \cite{Bengio-2000, khurana2023natural}. The development of deep learning models including Recurrent Neural Networks (RNNs), Recurrent Neural Network Language Model (RNNLM), Latent Semantic Analysis (LSA) \cite{landauer1998introduction}, Long Short-Term Memory (LSTM) \cite{hochreiter1997long}, and Gated Recurrent Units (GRUs) \cite{cho2014learning} is pivotal in harnessing the power of deep neural networks for comprehending and processing natural language.

The word embedding technique introduced by \cite{mikolov2013distributed} encodes words within sentences as dense vectors in a continuous space, capturing semantic relationships between words and ensuring that words with similar meanings have similar representations. This breakthrough facilitates efficient information retrieval within sentences and enhances the ability to predict subsequent words, thereby advancing language models. While deep learning models can enhance accuracy with ample training data, they initially struggled to map sequences in language processing. Sutskever et al.  \cite{sutskever2014sequence} proposed an encoder-decoder model architecture, leveraging RNNs, to effectively map input sequences to corresponding outputs, marking a significant advancement in NLP architecture. Despite its strengths, this model faces challenges in handling longer sentences with varying input lengths, a limitation addressed by the Transformer architecture.

The advent of Attention mechanism 
enabled neural networks to selectively focus on sentence elements, enhancing contextual understanding and handling longer sequences \cite{vaswani2017attention}. The self-attention mechanism enables the model to focus on various segments within sentences, regardless of their proximity, assigning weights based on their relevance. Among the breakthrough applications stemming from this advancement is ChatGPT, a generative AI Large Language Model (LLM)  trained on extensive text data that demonstrates the capacity for intelligent, human-like conversation across a spectrum of tasks. 
In less than nine months since its launch in Nov. 2022, ChatGPT has reached an impressive milestone with 100 million active monthly users and 1.6 billion monthly visitors across diverse domains, marking a record-breaking growth in technological history \cite{chatGPT-statistics-2023}.

ChatGPT rapidly found a place in academia, intriguing students to utilize it for generating reports, essays, and code, and aiding in exam preparation, often outperforming humans. This surge sparked discussions on its impact on higher education. Recent studies 
explore its potential to enrich teaching programs, expand learning materials, and enhance student assessment \cite{kasneci2023chatgpt, firat2023chatgpt, ellis2023new}. Furthermore, ChatGPT assists students in refining critical-thinking skills, test preparation, and real-time feedback, as seen in \cite{Gilson2022-medical-exam} where it achieved comparable results to a third-year medical student in medical examinations. 

Nevertheless, the application of ChatGPT in education has raised concerns about academic integrity, that students may submit generated work without effort, prompting calls for improved detection methods \cite{cotton2023chatting}. Scholars advocate updating academic integrity policies to address technology-driven plagiarism and suggest adapting examinations to emphasize higher-order reasoning  \cite{susnjak2022chatgpt, 2023ChatGPT-Rudolph}. Some express concerns about excessive reliance on ChatGPT, potentially leading to a decline in critical thinking skills \cite{Rahman-chatgpt-2023}. Moreover, biases and inaccuracies in ChatGPT's outputs are highlighted as its challenges in education \cite{DWIVEDI2023102642}. Despite notable performance, investigations reveal limitations in ChatGPT's mathematical capabilities, often falling short of graduate-level proficiency \cite{frieder2023mathematical}.

Debates persist over whether generative AI, like ChatGPT, will replace jobs and disrupt economies. McKinsey's recent research \cite{Mckinsey-report-generativeAI-2023} suggests viewing generative AI as a productivity enhancer rather than a job eliminator. They project a potential 3\% to 5\% annual productivity increase in the US through 2030 by combining generative AI with existing advanced processes like automation.
Their report indicates that leveraging generative AI, such as ChatGPT, allows the workforce to redirect efforts from mundane tasks to more creative and collaborative endeavors. To prepare for this shift, higher education must adapt its teaching methods. Recent studies showcase ChatGPT's positive impact in education, spanning computer science, engineering, statistics, student perspectives, programming training, and responsible use considerations \cite{ellis2023new, hassan2023chatgpt,zheng2023chatgpt, YILMAZ2023100147, qadir2023engineering,hassani2023role}. However, given its emergence and challenges, further research is essential to delve deeper into the application of ChatGPT in CSE education.

\begin{figure*}[!t]
\centering
\vspace{2ex}
\includegraphics[width=.67\linewidth]{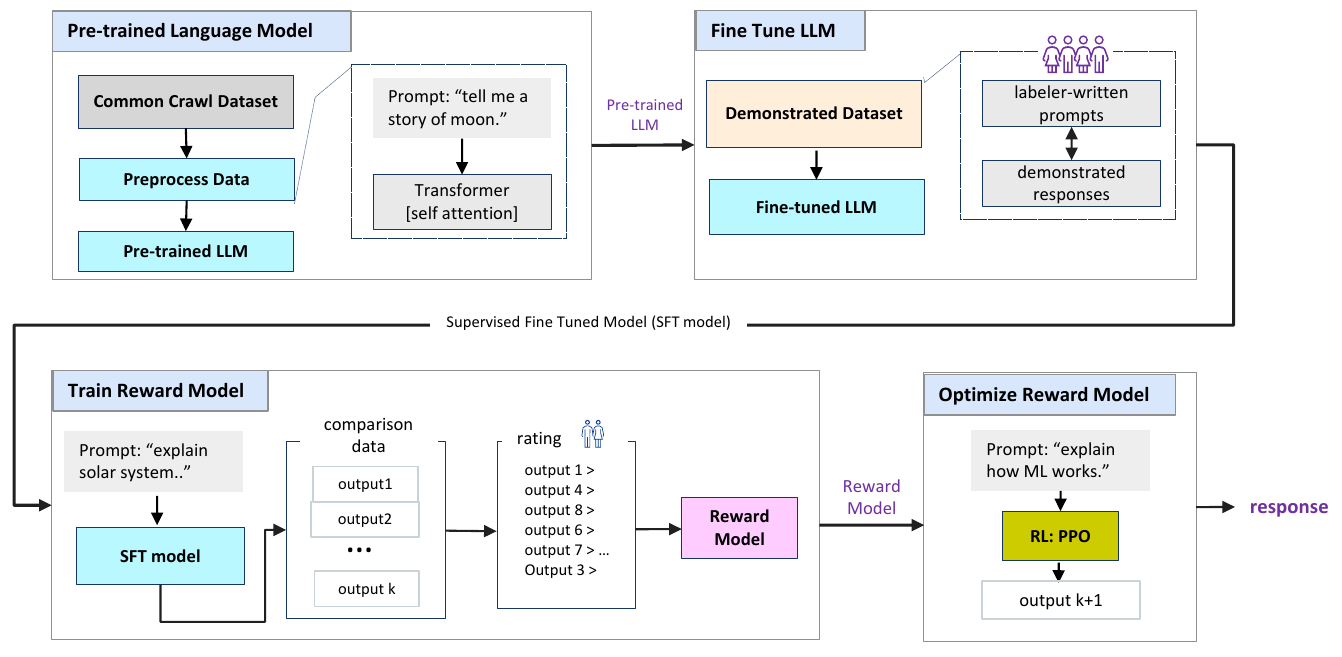}
\caption{Overall workflow of ChatGPT}
\label{fig:chatgpt-diagram}
\end{figure*}

In this paper, we focus on comprehensively analyzing ChatGPT's pedagogical potential within computer science and engineering education. We undertake a systematic approach, creating diverse educational practice problems while centering on pertinent areas like data science, programming, AI, machine learning, networks, and computer architecture. Through rigorous experimentation, we identify question types that challenge ChatGPT, offering valuable insights into its limitations. Notably, our study excludes less challenging queries and instead emphasizes personalized, project-oriented tasks, incorporating non-public datasets for evaluation. We introduce a comprehensive five-factor reliability analysis methodology, aiming to pinpoint ChatGPT's strengths and weaknesses. Our correlation analysis unveils intricate relationships between subjects, task types, and reliability factors: correctness, usefulness, clarity, coherence, and completeness, presenting actionable recommendations to augment ChatGPT's efficacy in education.

Subsequently, we conduct an in-depth investigation into each unresolved prompt, examining causal factors based on subject and task type. We offer suggestions 
on utilizing ChatGPT's flawed responses to assist enhanced learning as well as improving ChatGPT's performance in CSE education. Finally, we evaluate ChatGPT's cognitive capabilities, 
offering an assessment of its intellectual quality. The goal is to provide educators and learners with insights into what aspects of ChatGPT's responses can be trusted and where caution is warranted, which can guide the enhancement of learning objectives while establishing it as an effective educational tool. Embracing the inevitability of ChatGPT's societal impact, we suggest higher education's adaptation to propose tailored policies and methods, keeping up with the ever-increasing expansion of AI-assisted tools in educational settings.

The remainder of this paper is organized as follows.   Section \ref{sec:chatGPT} outlines ChatGPT's background, its architecture, and its educational applications. Following that, Section \ref{framework} introduces our proposed methodology, while Section \ref{sec:results} delves into our evaluation results, analysis,  recommendations, and ChatGPT's impact on learning and assessment in CSE education.  
Lastly, Section \ref{sec:con} concludes our study.

\section{ChatGPT: Background and Related Work}
\label{sec:chatGPT}
In this section, we provide an overview of the context and previous research work related to ChatGPT for education. The objective is to elucidate the foundational aspects and pertinent relevant literature, thereby laying the groundwork for further discussions on the topic and implying necessity of proposing a comprehensive analysis of the pedagogical potential of ChatGPT in computer science and engineering education.

\subsection{ChatGPT Overview and Architecture}

ChatGPT, part of OpenAI's GPT models, is an advanced conversational AI developed to produce human-like text responses and engage in dialogue. It has versatile applications, serving as virtual assistants, customer support chatbots, and an educational tool. Evolving from GPT-1 to GPT-3, these predecessors laid the foundation for ChatGPT. While GPT-4 introduces enhanced capabilities like image recognition, GPT-3.5, chosen for its accessibility, remains pertinent, especially for educators and student learners seeking optimal performance without a monthly subscription.

At its core, ChatGPT relies on GPT-3.5, a Large Language Model (LLM) using the Transformer architecture introduced by Vaswani et al \cite{vaswani2017attention}, in particular, using sparse attention in each transformer layer \cite{child2019generating}. This architecture efficiently handles data sequences. GPT-3.5 undergoes two primary stages: Pre-training and Fine-tuning. Pre-training involves learning statistical patterns and linguistic structures from a vast internet text corpus to predict subsequent words. With 175 billion parameters across 96 layers, GPT-3.5 stands as one of the largest deep learning models. Fine-tuning customizes it for diverse applications, including enhancing conversational abilities through training on specific datasets.

Figure \ref{fig:chatgpt-diagram} demonstrates the workflow ChatGPT. ChatGPT3 employs a pre-trained large language model \cite{brown2020language} trained over the Common Crawl dataset \cite{raffel2023exploring}. During its training process, it first preprocess data from the Common Crawl dataset,
data quality is refined by eliminating irrelevant or noisy information and aligning it with the model's needs. 
Additional methods include removing unwanted characters, correcting spellings, and fixing grammatical errors.

The backbone of the ChatGPT's pre-trained language model is using Transformer architecture, 
which employs self-attention mechanisms to weigh word importance, understand text context and dependencies. It contains a stack of encoder layers, a stack of decoder layers, and an output layer to generate the response from the prompt. Before the encoder and decoder layers, it first feeds the input data to word embeddings and position encoding. Transformer architecture enables the computation of the relationship among words in the input sequence and allows the model to focus on those words that are related to the current word. 

Subsequently, ChatGPT utilizes supervised learning and human feedback demonstration data to fine-tune the pre-trained LLM model. This heavily involves human experts and feedback. The demonstration dataset contains example prompts and expected responses created by these trained human experts. Then, it retrains/fine-tunes the LLM model to enhance the model performance, ensuring its adaptability and versatility across different applications.
In the next stage, 
a reward model is trained to guide the model to produce conversational-style responses according to the best-rated output. 

First, for each prompt, it collects comparison data which uses the prompt as input in the language model to produce 4 to 9 outputs/responses \cite{brown2020language, stiennon2022learning}. Then, the trained human experts rate each response from the best to the worst. This comparison data is used to train a reward model, which is the same language model without the last output layer so that it produces a scalar reward value for each ranked output. The reward model is trained to predict the response that human experts consider the best when given a prompt, using the training comparison dataset as a basis. 

As the last stage of optimizing the reward model, new prompts are used to output rewards from the reward model, which are subsequently utilized to train a Proximal Policy Optimization (PPO) \cite{schulman2017proximal} based on reinforcement learning (RL). The environment operates within a bandit framework, simulating a reinforcement learning environment. It generates random prompts and seeks human-preferred responses guided by the reward policy. Upon receiving the prompt and response, the reward model generates a reward. This refined policy aims to generate the most fitting responses for each input prompt \cite{stiennon2022learning}. Along with more queries with ChatGPT, the reward model is continuously optimized to better align with human preferences. This optimization process follows a framework 
 which delineates the conditions under which a model is aligned with user intent \cite{askell2021general}.

\subsection{Related Work on ChatGPT for Education}
Table \ref{tab:paper-review} provides a comprehensive summary of recent research efforts 
delving into the evaluation of ChatGPT's potential in the field of education. 
Hassani et al. \cite{hassani2023role} conduct a comparative analysis of language processing models, focusing on ChatGPT's role in data science education. Highlighting its potential to advance workflows and outcomes for data scientists, the study acknowledges ChatGPT's imperfections. It notes that, like any language model, ChatGPT is not perfect, and its accuracy depends on various factors, including the quality and diversity of training data, the complexity of input text, and the nature of the task.  
The work in  \cite{hassan2023chatgpt} showcases ChatGPT's application as a personalized data scientist. 
The system, VIDS (Virtual Interactive Data Scientist), functions as an AutoML assistant. 
Despite improved precision, the model occasionally generated 
displayed shortcomings, especially when presented with few-shot learning examples.

Ellis et al. \cite{ellis2023new} discuss ChatGPT excels with tailored prompts, outperforming conventional search engines, especially in addressing nuanced statistical queries. However, ChatGPT generates varied 
responses to prompts about confidence intervals and p-values. Also, instead of banning ChatGPT, the paper recommends educators adopt alternative approaches for students to harness generative AI effectively. 
Examining ChatGPT's role in data science education, 
the work in \cite{zheng2023chatgpt} acknowledges its efficacy as an educational tool. However, limitations arise in its effectiveness for assessing problem-solving questions with multiple correct answers. The study included 28 students in its exploration. 
Student feedback showcased positive experiences but highlighted challenges in critical thinking and problem-solving support.

\begin{table*}[!t]\scriptsize
\caption{Summary of recent studies on evaluating the impact of ChatGPT in education}
\vspace{-1ex}
\label{tab:paper-review}
\centering
\scalebox{0.85}[0.85]{
\begin{tabular}{|p{2em}|p{14em}|p{50em}|p{12em}|}
\hline
\multicolumn{1}{|c|}{\textbf{Research}} &
  \multicolumn{1}{|c|}{\textbf{Target Field}} &
  \multicolumn{1}{|c|}{\textbf{Major Statements}} &
  \multicolumn{1}{|c|}{\textbf{Capabilities Examined}} \\ \hline
\cite{hassani2023role} &
  Data Science; Academic Integrity &
 ChatGPT will become more widely used and accepted in the data science field, helping to improve workflows and deliver better results. The accuracy and dependability of ChatGPT's responses depend on factors such as training data quality, diversity, input text, and type of questions.  
&
  Theoretical question, code debugging\\ \hline
\cite{hassan2023chatgpt} &
  Computer Science and Engineering &
  Introduced the concepts of global micro-agents, creating a structure to maintain a cohesive conversation. Employing targeted prompts improves precision and control, rather than relying on a single overarching prompt. 
  &
  Conversation, data set summarize, task manager, micro-operation, \\ \hline
\cite{banerjee2023understanding} &
   Computer Science and Engineering &
   ChatGPT or equivalent AI models will eventually become part of our daily lives, and it's pivotal to exploit their fullest potential in existing education processes, such as teaching, learning, and assessment.   &
  Programming topics, Electrical Engineering assignments\\ \hline
\cite{zheng2023chatgpt} &
  Data analytic; Data science &
  In Data Science education, ChatGPT struggles with assessing problem-solving questions since there could be multiple available and correct answers, rather than unique ones. &
  N/A \\ \hline
\cite{qadir2023engineering} &
  Education; Engineering &
The field of engineering education will inevitably adopt tools such as ChatGPT and equivalent model, given their convenience and significant potential for assistance. It is crucial to reconsider our evaluation methods, to prevent unethical misuse while upholding the productivity benefits. &
   Code generation, conceptual questions, math questions\\ \hline
 \cite{feng2023investigating}&
   Computer Science; User sentiment &
   Examined the sentiment of ChatGPT regarding code generation across platforms such as Reddit and Twitter. & Sentiment analysis, code generation
\\ \hline
\cite{ellis2023new} &
  Education; Statistics &
  ChatGPT excels when given a specific prompt that users can refine as needed. In contrast, traditional search engines may struggle with overly specific or verbose input. &
  N/A \\ \hline
\cite{singh2023exploring} &
  Computer Science &
  The survey suggests that students find AI tools valuable for obtaining explanations on various topics, and there is a consensus that these tools should be permitted in education. However, it highlights the need for a redesign in assessment methods and guidelines to use the tool for learning effectively.&
  Application development \\ \hline
\cite{malinka2023educational} &
  Education; Academic Integrity &
 The misuse of ChatGPT in lower-division courses and foundational concepts can have adverse effects on learning, potentially impeding students' progress as they advance to more challenging topics. 
&
  Fulltext Exam, Term essays,   programming assignment \\ \hline
\cite{qureshi2023exploring} &
  Computer Science &
 Students who are permitted to use ChatGPT for solving programming problems demonstrate better performance compared to the group that did not. &
  Programming assignment \\ \hline
 \cite{Rahman-chatgpt-2023}& 
 Computer Science & 
The survey among teachers suggested that educators believe ChatGPT is a powerful assisting tool, however, it has not gained enough trust in assessing student assignments. &Lesson planning, code review/debugging, generating code solutions, test questions.\\ \hline 
 \cite{frieder2023mathematical}& Mathematics& 
 Findings show that ChatGPT struggles with providing high-quality proofs and consistence calculations. & Grad-Text, Olympiad-Problem Solving, Math (Algebra-Precalculus)\\ \hline 
 \cite{opara2023chatgpt}& Computer Science; Academic Integrity & 
Suggested that the model tends to be verbose and overly relies on specific terms. They propose that these issues arise due to biases in the training data, where a preference for length may overshadow the need for a comprehensive response.  &N/A\\ \hline 
 \cite{farrokhnia2023swot}& Education& 
Conducted SWOT analysis of ChatGPT in education. Suggested adjustments to curricula and emphasized the need for further empirical research to optimize ChatGPT's educational applications. &Strengths, weakness, and opportunities in personalized learning\\ \hline
 \cite{YILMAZ2023100147} & Computer Science & The study involving 45 undergraduate programming students found that incorporating ChatGPT can enhance creativity,  thinking, cooperation, problem-solving, algorithmic and critical thinking compared to a control group.  & Programming application; critical thinking; problem solving \\ \hline
 \cite{susnjak2022chatgpt}&Academic Integrity& ChatGPT poses a significant threat to online examinations, as the results reveal its advanced critical thinking capabilities with minimal input prompts, enabling potential cheating by students. &Critical thinking\\ \hline
 \cite{kasneci2023chatgpt}&Opportunities and Challenges in Education&Addressed key challenges such as copyright issues, bias, overuse,  hard-to-identify fact and fiction text, etc. Additionally, provides recommendations on effectively addressing these challenges to ensure responsible and ethical usage in education.&N/A \\ \hline
\cite{Bob-or-bot}&Computer Science, AI-content Detection& This study demonstrated that ChatGPT can achieve passing grades at different levels of study with differing assessment models in introductory CS courses. However, ChatGPT does not perform better than an ordinary CS student. There's a high likelihood that students using ChatGPT material alongside their own responses could achieve higher grades in assessments. 
& Academic performance and plagiarism detection tools. 
\\ \hline

\end{tabular}
}
\end{table*}

The research in \cite{qureshi2023exploring} 
 assessed the effectiveness of using ChatGPT for solving programming problems, involving 24 students. Results showed that the group utilizing ChatGPT performed better, achieving higher scores in less time, despite facing code inaccuracies. 
The study in \cite{malinka2023educational} raises concerns about ChatGPT's overuse in lower division courses, potentially impeding student learning and graduation rates. 
The authors evaluate ChatGPT's abilities in solving assignments in computer security specialization. 

Singh et al. \cite{singh2023exploring} echo similar concerns in their exploration of ChatGPT's challenges and potential risks in higher education. Surveying 430 students, the paper reveals students' apprehensions about misuse despite their familiarity with ChatGPT. While acknowledging its benefits in writing and code generation, students express concerns about their limited understanding of the tool.
The paper advocates for integrating ChatGPT in education with defined guidelines instead of outright restrictions. 

The study in \cite{qadir2023engineering} explores ChatGPT's potential applications in engineering education while highlighting its imperfections, including the generation of potentially biased or incorrect information. The article takes an unconventional approach, engaging ChatGPT to differentiate acceptable from non-acceptable educational use, addressing concerns about plagiarism, academic integrity, and its impact on online exams.

The work in \cite{frieder2023mathematical} presents a testing methodology 
to gauge ChatGPT's math capabilities. Establishing a benchmark for large language models, the paper explores ChatGPT's practicality in mathematical contexts. It reveals ChatGPT 
lack the adequate proficiency for advanced university-level math due to limitations in delivering high-quality proofs and calculations. However, it underscores ChatGPT's potential as a valuable assistant, especially with users able to assess its output.

Firat et al. \cite{firat2023chatgpt} explored scholars'
views on integrating ChatGPT and AI in universities. Findings from 21 participants echoed existing literature, highlighting AI's potential benefits and challenges in education. 
However, concerns arose about assessment methods and ethical implications. The study emphasizes the necessity for clear guidelines and policies to prevent misuse and educate students on effectively integrating AI into their learning processes.
Moreover, the study in \cite{opara2023chatgpt} assesses ChatGPT in education and research, acknowledging its fast conversational responses but noting limitations like the absence of citations and possible inaccuracies. It emphasizes concerns about hindering learners' creativity and the model's restricted scope, urging action to address these issues.

\begin{figure*}[!t]
\centering
\includegraphics[width=17cm]{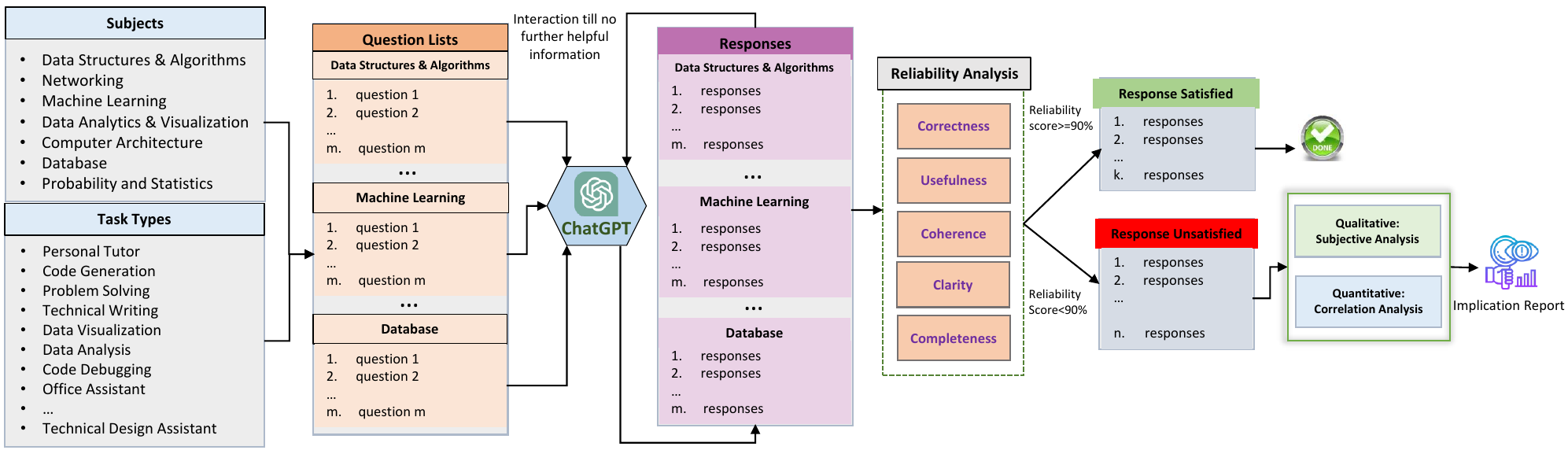}
\caption{Overview of the proposed methodology for analyzing the potential of ChatGPT in Computer Science and Engineering education}
\label{fig:overall-method}
\end{figure*}

Rahman et al. \cite{Rahman-chatgpt-2023} delve into ChatGPT's role in programming education. 
It includes coding experiments like code generation, pseudocode creation, and correction, validated via an online judge system. Additionally, a survey with students and teachers evaluates ChatGPT's impact on programming education. Results show around 50\% of teachers rated ChatGPT 3/5, indicating partial trust in programming education. 
Farrokhnia et al. \cite{farrokhnia2023swot} conducted a SWOT analysis of ChatGPT in education. It outlined strengths like generating personalized responses and potential for improvement, along with opportunities for personalized learning. However, it noted weaknesses such as limited understanding and biases, highlighting threats to academic integrity and cognitive skills. 

The paper in \cite{feng2023investigating} 
explores ChatGPT's code generation capabilities across different programming languages and analyzes sentiment regarding ChatGPT on social media. 
The finding reveals that Python and JavaScript are the most popular languages, 
with ChatGPT used for various purposes (e.g., debugging, interview preparation).
In \cite{YILMAZ2023100147}, 45 undergraduates split into two groups—one using ChatGPT during programming practices—showed improvements in critical thinking and problem-solving. However, motivation for tough tasks didn't change, suggesting exploring new ways to boost motivation during challenging assignments.

The study outlined in \cite{Bob-or-bot} devised a dual-anonymous protocol to evaluate responses to questions from four end-of-module assessments in a basic computer science course. These responses, from both ChatGPT and students, formed part of a quality assurance investigation. Findings revealed ChatGPT achieved a pass rate exceeding 40\%, notably scoring over 85\% in the introductory module but falling below 40\% in questions relating to personal development planning and reflection. Additionally, the study assessed the effectiveness of plagiarism detection tools such as GPT-2 and TurnItIn AI in identifying AI-generated content. Results demonstrated TurnItIn AI detected all AI-generated work with 100\% accuracy, while GPT-2 had partial success.

The work in \cite{Nikolic2023_ChatGPT_engineering} explored ChatGPT's impact on engineering education across seven Australian universities, assessing its effectiveness in ten subjects. Tasks encompassed online quizzes, oral assessments, and coding exercises. Results, akin to student grading, indicated ChatGPT's success in three subjects, failure in five, and indeterminate outcomes in two. Variability in performance across subjects and task types was noted, with suggestions offered for maximizing strengths and mitigating weaknesses to enhance engineering learning.

\section{Proposed Methodology}
\label{framework}

This section outlines the proposed methodology for analyzing the potential of ChatGPT for CSE education. Initially, we introduce the overarching methodology, followed by an exposition of the five factors contributing to reliability scores and the metric formula. Subsequently, we present a correlation analysis-based approach to explore causal factors, complemented by qualitative analysis aimed at providing deeper insights into the quantitative results. Additionally, we present an examination of implications and weaknesses for comprehensive analysis.

\subsection{Overview of Methodology}

Figure \ref{fig:overall-method} demonstrates the 
comprehensive methodology proposed in this work for assessing the potential of ChatGPT in supporting computer science and engineering education. Initially, we 
specify various 
primary subjects within undergraduate CSE curricula at the University level including 
data structures \& algorithms, networking, machine learning, data analytics \& visualization, computer architecture \& organization, database, 
and probability \& statistics. Subsequently, we summarized twelve common task types in Table \ref{tab:task-type}, which serve as a foundation 
for crafting prompts to engage with ChatGPT. 

Acknowledging ChatGPT's proficiency in certain tasks, our prompt design strategy centers on formulating challenging yet college-level questions pertinent to the subject matter and task types. Employing an iterative approach, we systematically generate five questions per subject across various task types (35 scenarios in each iteration), aimed at pushing the boundaries of ChatGPT's response capabilities. Through successive interactions, we gather responses and continually refine the prompts until we accumulate at least ten unsatisfactory answers. Following this data collection phase, reliability analysis is conducted on all responses, utilizing the evaluation metrics and formula described in Subsection \ref{subsec:reliability}. Responses yielding a reliability score exceeding a threshold of 90\% are deemed satisfactory. For those falling below this threshold, a detailed quantitative and qualitative analysis of causal factors is undertaken to delve into the shortcomings.

\begin{table*}[!t] \scriptsize
\caption{Task types considered for analyzing potential of ChatGPT}
\vspace{-1ex}
\label{tab:task-type}
\centering
\scalebox{0.92}[0.92]{
\begin{tabular}{|l|l|l|}
\hline
\multicolumn{1}{|c|}{\textbf{\#}} & \multicolumn{1}{c|}{\textbf{Task Type}} & \multicolumn{1}{c|}{\textbf{Description}} \\ \hline
1  & Conceptual question          & Understanding and explaining concepts and knowledge.                         
\\ \hline
2  & Idea generation              & Suggestions on what and how to approach a particular problem.                   
\\ \hline
3  & Office assistant             & Creating a PPT presentation and tables from a given topic. 
\\ \hline
4  & Technical design   assistant & Assisting in developing architectures, workflows, and processes.     
\\ \hline
5  & Writing technical   report   & Drafting technical reports within the field of CSE.        \\ \hline
6  & Personal tutor         & Providing tutoring services and crafting personalized learning plans.     
\\ \hline
7  & Problem solving         & solving math and algorithm problems.                          \\ \hline
8  & Reasoning and   analysis & Analytical reasoning, inference, and mathematical analysis.   \\ \hline
9  & Code generation              & Generating code and designing pipelines for software development.                   
\\ \hline
10 & Code debugging               & Finding errors and bugs in   codes.                   \\ \hline
11 & Data analysis                & Analyzing data to provide   insights and findings.      \\ \hline
12 & Data visualization           & Plotting charts and graphs.                         \\ \hline
\end{tabular}
}
\end{table*}

From a quantitative perspective, 
the reliability analysis involves 
examining the subjects, task types, and five key reliability factors: correctness, usefulness, clarity, coherence, and completeness. A lower reliability score might stem from task type, subject matter, or their interplay, significantly influencing correctness, usefulness, completeness, and beyond. 
The objective is to discern how subjects as well as the educational task types, impact a user's perception of ChatGPT's reliability. 
Employing quantitative correlation analysis provides a robust assessment of ChatGPT's suitability as an educational tool, offering insights crucial for enhancing its reliability, and facilitating its use for CSE education. Details on analysis and the implications for improvement are presented in Section \ref{subsec:analysis}.

In addition to the quantitative analysis, we 
delve into our experience with unsatisfactory prompts to elucidate  
the quantitative findings. This approach allows us to offer valuable insights, provide intricate details, and meticulously evaluate ChatGPT's limitations in CSE education. We curate screenshots of unsuccessful prompts, conducting a granular analysis to pinpoint specific issues encountered line by line. This examination enables us to 
offer recommendations for ChatGPT's enhancement, envisioning how it can mitigate these identified limitations. Further elaboration on these observations and improvement suggestions is presented in Section \ref{subsec:examples}.

\subsection{Reliability Factors and Evaluation Metrics} \label{subsec:reliability}

We 
assess the reliability of all ChatGPT responses using a framework comprising five key factors. Each factor is evaluated on a scale from 1 to 5, and the cumulative scores yield the overall reliability score for each response. Responses achieving a reliability score of 90\% or higher are considered acceptable. However, for responses falling below this threshold, a comprehensive analysis follows, integrating quantitative and qualitative methods. This examination aims to uncover the reasons behind their lower reliability, shedding light on the primary factors impacting response quality. The reliability factors are described below: 

\noindent  
\textit{- Correctness:} This metric assesses the accuracy of the generated answers, focusing on factual correctness, coherence, and absence of synthetic content. Additionally, it considers the accuracy and clarity of visual aids like diagrams or plots when presenting factual information. 

\begin{itemize}[leftmargin=*]
\small{
    \item 1- Incorrect: Generated responses lack factual evidence and information might be fabricated. 
    \item 2- Contains a mix of accurate and inaccurate ($>$50\%) information.
    \item 3- Contains a mix of accurate ($>$60\%) and inaccurate information. 
    \item 4- Mix of accurate ($>$80\%) and inaccurate ($<$20\%) information. 
    \item 5- Fully Correct: Generated responses are factually sound.
    }
\end{itemize}

\noindent
\textit{- Usefulness:} This metric evaluates the quality or fact of being useful from a user's perspectives. It gauges practicality, determining whether the answer fulfills the user's needs. 
We can ask questions when evaluating this factor, such as "Would I use this answer?" or "How much of them would I use?". 
\begin{itemize}[leftmargin=*]
\small{
\item 1- Useless: Generated responses lack practical usability. 

\item 2- $<$40\% of usefulness, meaning user has to add another 60\% more work to make the response useful.

\item 3- $<$60\% of usefulness, meaning user has to add another 40\% more work to make the response useful.

\item 4- $>$80\% of usefulness, meaning user need to add another 20\% more work to make the response useful.
\item 5- Useful: Generated responses are useful (near 100\%, repeated content is acceptable), that user can use it without any further effort and major adjustment.
}
\end{itemize}

\noindent
\textit{- Clarity:} This metric evaluates how well the language model can produce output that is fluent and easily comprehensible 
and offers sufficient 
evidence to support the response. 
The output should demonstrate proper grammar, use 
appropriate language, while avoiding  unnecessary or confusing information.

\begin{itemize}[leftmargin=*]
\small{
\item 1- Hard to Understand: Generated responses are difficult to understand maybe due to lack of context and evidence, or the output does not represent natural language, or poor or incorrect grammar. 
\item 2- Contains a mix of clarified and unclarified ($>$50\%) information. 
\item 3- Mix of clarified ($>$60\%) and unclarified ($<$40\%) information. 
\item 4- Mix of clarified ($>$80\%) and unclarified ($<$20\%) information. 
\item 5- Easy to Understand: Responses are easy to understand; output is well-constructed 
with 
adequate evidence and context.
}
\end{itemize}

 \noindent
\textit{- Coherence:} This metric assesses the model's ability to maintain a coherent conversation and provide non-repetitive, context-aware responses. If the user requires further insights, the model should seamlessly continue the conversation, offering valuable information without unnecessary duplication. 
\begin{itemize}[leftmargin=*]
\small{
\item 1- Incoherent: Generated responses lack cohesiveness from the previous response; the responses either are repetitive or lack awareness of the conversation context.  
\item 2- Contains a mix of coherent and incoherent ($>$50\%) information. 
\item 3- Mix of coherent ($>$60\%) and incoherent ($<$20\%) information.
\item 4- Mix of coherent ($>$80\%) and incoherent ($<$20\%) information.
\item 5- Coherent: Responses show coherence with prior answers, minimal repetition, and an understanding of the conversation's context.
} 
\end{itemize}

\noindent

 \noindent
\textit{- Completeness:}  
Given that a user can engage with ChatGPT incrementally, it assesses the entirety of responses, ensuring they are comprehensive 
across all interactions.

\begin{itemize}[leftmargin=*]
\small{
\item 1- Incomplete: Generated responses are not 
complete and the prompt is partially answered. It falls within this range if the answer's completeness is below 20\% (over 80\% incompleteness). 

\item 2- $>$60\% incomplete, needs 60\% more work to make it complete.

\item 3- $>$40\% incomplete, needs 40\% more work to make it complete.

\item 4- $>$80\% complete, needs 20\% more work to make it complete.
\item 5- Complete: Responses are fully complete (e.g. code is fully generated and libraries are properly imported, so that the code can be run without further modification).
}
\end{itemize}

\noindent
\textit{- Weights Assigned to Each Metric:} Acknowledging the varying significance of evaluation metrics, we have assigned specific weights to each metric, aligning them with their relative importance about the examined course subjects and tasks. 
The detailed distribution of these weights is outlined in Table \ref{tab:weight}, reflecting their respective significance in guiding educators' decision-making processes. 
The allocation of weights 
is derived from our analysis encompassing diverse subjects, tailored to assist educators and learners. Notably these weights, while structured for this context, may differ in other  domains or educational settings. They are not fixed and could be adjusted to accommodate the unique requirements 
of different subject areas. The flexibility of these weights ensures adaptability across diverse educational environments.

\begin{table}[!t] \footnotesize
\caption{Weight of each reliability factor}
\vspace{-1ex}
\label{tab:weight}
\centering
\begin{tabular}{|l|l|}
\hline
Metric & Weight percentage \\ \hline
Correctness & 40\% \\ \hline
Usefulness & 20\% \\ \hline
Clarity & 20\% \\ \hline
Coherence & 10\% \\ \hline
Completeness & 10\% \\ \hline
\end{tabular}
\vspace{-1ex}
\end{table}

The Reliability Score in our analysis is calculated as below:

\vspace{0.8ex}
\noindent
$\textit{R-Score} = (\textit{Correctness} \times 0.4) + (\textit{Usefulness} \times 0.2) + (\textit{Clarity} \times 0.2) + (\textit{Coherence} \times 0.1) + (\textit{Completeness} \times 0.1)$
\vspace{0.8ex}

The weighted R-Score is a composite measure derived from evaluating various metrics—correctness, usefulness, clarity, coherence, and completeness. Each metric's contribution to the overall score is determined by its assigned weight. The proposed equation aggregates the weighted metrics to provide a comprehensive assessment of the reliability of ChatGPT's responses in CSE educational contexts.

\vspace{-1.3ex}
\subsection{Integrating ChatGPT in Learning and Assessment}
\vspace{-0.2ex}
Anderson et al. \cite{anderson2001taxonomy} redefined Bloom's taxonomy by 
emphasizing the hierarchical nature of learning objectives. These objectives encompass various levels of skills development, including remembering, understanding, applying, analyzing, evaluating, and creating (presented in Subsection \ref{sec:Bloom}). Notably, these skills are not necessarily acquired sequentially but represent a spectrum illustrating the depth and extent of skill development throughout education.
The conducted reliability analysis in this work 
aligns with Bloom's taxonomy in evaluating ChatGPT's proficiency across various cognitive skill levels. 
Correctness pertains to the foundational level of remembering and understanding, ensuring factual accuracy and comprehension of content. Usefulness and clarity correspond to higher levels such as applying and analyzing, emphasizing the practical application and critical evaluation of information. 

Coherence and completeness signify the synthesis and creation of new knowledge, aligning with evaluation and creation levels in Bloom's taxonomy. Thus, the proposed reliability analysis serves as a complementary assessment, mapping ChatGPT's performance onto Bloom's taxonomy, enabling a comprehensive understanding of its effectiveness across diverse cognitive skill levels crucial for educational integration (described in Subsection \ref{sec:implication}). 
Particularly, Bloom's taxonomy serves as a framework to assess ChatGPT's proficiency across essential skill metrics, rating them on 
a scale of excellent, good, fair, and poor. This evaluation offers insights into ChatGPT's capabilities aligned with the spectrum of cognitive skills outlined in Bloom's taxonomy. Moreover, it informs suggested implications for higher education, spanning areas such as 
open challenges of reliability, fairness, and integrity. Hence, CSE programs are urged to update educational policy, curriculum, and assessment methods to adapt to such challenges.

\begin{table*}[!t]\scriptsize 
\caption{Selected unsatisfactory prompts that challenged the effectiveness of ChatGPT and their reliability scores }
\vspace{-1ex}
\label{tab:reliability-score}
\centering
\scalebox{0.82}[0.82]{
\begin{tabular}{|p{2.5em}|p{24em}|p{8em}|p{6em}|p{4em}|p{4.5em}|p{4em}|p{3em}|p{4em}|p{5em}|p{5em}|}
\hline
\textbf{Prompt \#} & \textbf{Prompt Summary} & \textbf{Subject} & \textbf{Task Type} & \textbf{R-Score} & \textbf{Correctness} & \textbf{Usefulness} & \textbf{Clarity} & \textbf{Coherence} & \textbf{Completeness} & \textbf{Better than Search?} \\ \hline
1 & Analyze uploaded malware and benign CSV files to plot T-distributed stochastic neighbor embedding in two-dimensional space. Give insight into the differences and similarities between malware and benign data.       &  Data Analytics \& Visualization      &  Data Visualization         & 86\%                    &   4          &     4       &      5   &    5       &    4          &   Yes                    \\ \hline
2 & Write a Python code that uses the library Networks to create a bipartite graph and create a function to find a maximum matching. &     Networking    &     Code Generation    &        78\%             &      4       &      5      &    3     &     3      &      4        &       Yes         \\ \hline
3 &   Binomial distribution with home security system requirement.    &   Probability \& Statistics     &    Personal Tutor       &   76\%                  &       2      &     5       &     5    &      5     &      5       &            Yes        \\ \hline
4 &   Applying Radix sort and calculate the bytes memory required.    &    Data Structure \& Algorithm    &      Personal Tutor     &    72\%                 &      2      &      5      &     5    &      5     &       3       &           No            \\ \hline
5 & Plot a clustered column chart with error bars across the column categories.       &  Data Analytics \& Visualization      & Data Visualization          &   86\%                  &  4           &   4         &   5      &    5       &       4       &   Yes                   \\ \hline
6  & Use the f1.xlsx file and give me some insights into obfuscated methods including 'null', 'shuf', 'split', and 'var' are downgraded for each ML compared with the ‘non\_obf’ column. & Data Analytics \& Visualization      &  Data Analysis      &     50\%               & 2          &  2        &     3  &    5    &     2         &     Yes                  \\ \hline

7  & How to create a train validation dataloader in PyTorch?   & Machine Learning       & Code Generation         &    52\%                 &       1      &      1     &      5   &      5     &     5         &  Yes                  \\ \hline
8  & Create an entity-relationship (ER) model for an online banking system.    & Database        &  Technical Creation         &  88\%                   &   5          & 3          & 5        &  5         &   3           &    No                  \\ \hline
9   &  Design a database system for a facial recognition application.   &      Database   &  Technical Creation         &  76\%                   &      5       &      1      &     5    &    5       &   1           &  No                 \\ \hline
10 & A 1\-inch\-diameter coin is thrown on a table covered with a grid of lines two inches apart. What is the probability the coin lands in a square without touching any of the lines of the grid?      &  Probability \& Statistics      & Problem Solving          &  52\%                   &    1         &    1        &   5      &       5    &    5          &    No                   \\ \hline
11& A Given an integer array nums, return all the triplets [nums[i], nums[j], nums[k]] such that i != j, i != k, and j != k, and nums[i] + nums[j] + nums[k] == 0.      &  Data Structure \& Algorithm      & Problem Solving          &  86\%                   &    4         &   4       &   5      &       4    &    5          &    No                   \\ \hline
12& What is 5ED4-07A4 when these values represent unsigned 16-bit hexadecimal numbers? The result should be written in hexadecimal. Show your work.    &  Computer Architecture    & Problem Solving          &  42\%                   &    1         &    1      &   3    &       4    &    5          &    No                   \\ \hline
13 & Calculate the binary representation of the decimal number (-20.5). Assume that the floating-point representation format is IEEE 754 single precision.       &  Computer Architecture       & Problem Solving          &   86\%                  &   4          &     4       &   5      &  4         &    5          &  Yes                    \\ \hline
14 & For the following MIPS language instruction first determine the instruction type. Next, convert it to the corresponding MIPS assembly language instruction: 0000 0010 1010 1111 0101 0000 0010 0000       &  Computer Architecture       & Problem Solving          &   78\%                  &   4          &     3       &   4      &  4         &    5         &  Yes                    \\ \hline
15 & Create PowerPoint slides explaining how Transformers work.       &  Machine Learning       & Office Assistant          &   78\%                  &   5          &     2       &   5      &  4         &    1          &  No                    \\ \hline
\end{tabular}
}
\end{table*}

\section{Results, Analysis, and Insights}
\label{sec:results}
This section delves into our experimental evaluation of ChatGPT's reliability in CSE education. We analyze unresolved prompts (detailed in Table \ref{tab:reliability-score}) encountered while interacting with ChatGPT. For example, when requesting code generation for a CNN model and a train-validation dataloader, errors were identified, leading to an unsatisfactory prompt (prompt \#7) within machine learning topics. 
We quantitatively examine reliability factors 
concerning subject, task type, and specific prompts. Unsatisfactory prompts prevail in all CSE subjects and task types (e.g. code generation, problem-solving, and data analysis). Specifically, as shown in Table \ref{tab:reliability-score}, computer architecture, data analysis, machine learning, and probability questions have notably lower reliability scores, reaching as low as 42\% (\#12), 50\% (\#6), 52\% (\#7) and 52\% (\#10), respectively.
Next, we perform an analysis of each unresolved prompt, examining causal factors based on subject and task type. We provide suggestions to address errors and enhance ChatGPT's performance. Additionally, we assess ChatGPT's cognitive abilities in CSE education to guide educators and learners on its reliability and limitations.

\subsection{Reliability Analysis}
\label{subsec:analysis}

\subsubsection{Reliability Scores}
Utilizing the reliability score formula described in Section \ref{subsec:reliability}, we calculate the R-Score for each prompt tested. We then determine its percentage score (R-Score/5) and present in Table \ref{tab:reliability-score} the prompts that did not meet the predefined reliability threshold (90\%).
\begin{figure}[!t]
\centering
\includegraphics[width=0.99\linewidth]{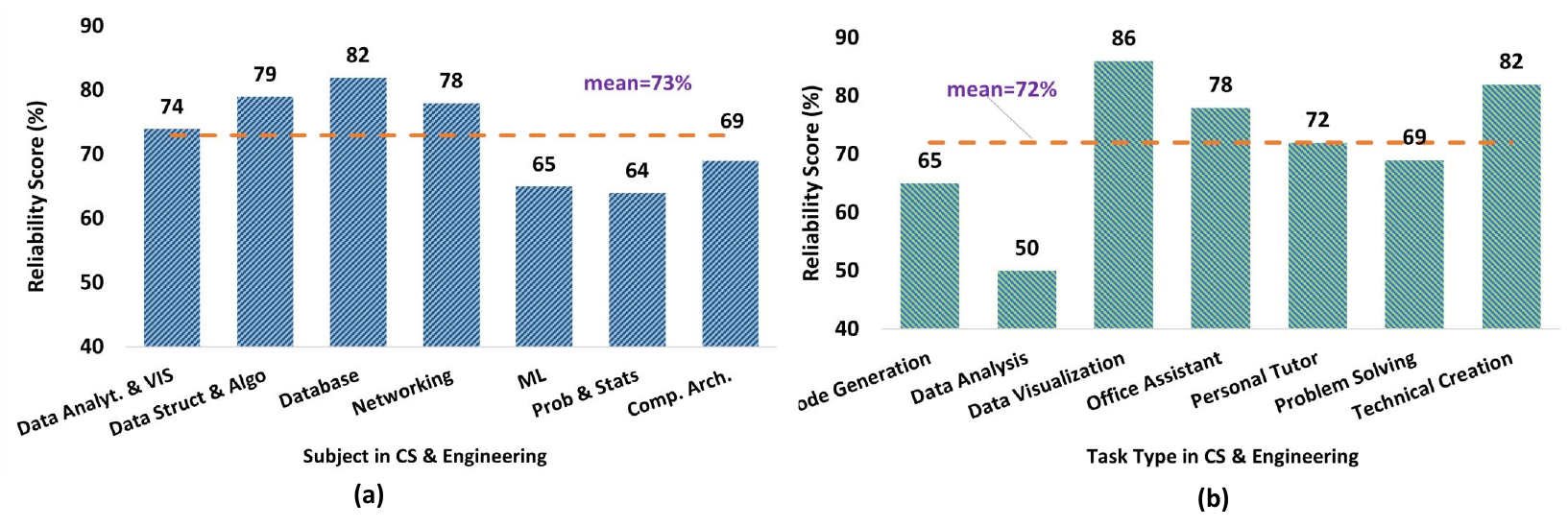}
\caption{Reliability scores (a) by subject, (b) by task type}
\label{fig:reliability-score}
\vspace{-1ex}
\end{figure}
Figure \ref{fig:reliability-score} depicts the summative reliability scores of the unsatisfactory prompts by subject and task type resulting from our experiments. Figure \ref{fig:reliability-score}-(a) outlines that among CSE subjects, different areas 
faced limitations, averaging a reliability score of 73\%. Notably, the subjects in machine learning, probability and statistics, and computer architecture and organization scored below this average. 

Figure \ref{fig:reliability-score}-(b) illustrates that challenging tasks averaged a 72\% score, with  data analysis, code generation, and problem-solving falling below this mark. While ChatGPT showcased proficiency in code generation, unreliability issues emerged upon closer inspection, such as incorrect code outputs. For instance, in prompt \#7, the generated code appeared accurate but produced incorrect results when run. Similarly, in prompt \#10, ChatGPT showed inconsistencies while solving probability questions, providing conflicting answers within the reasoning process. While competent in computer architecture, it struggled with fundamental concepts like binary number operations and converting MIPS instructions.  We observed that ChatGPT alternates between recalling memorized answers and attempting logical reasoning by running code to generate responses. This 
has led to inconsistencies and reliability issues when applying knowledge to practical scenarios, technical tasks, problem-solving, and advanced analyses.

\subsubsection{Reliability Factors Analysis}
\begin{figure}[!t]
\centering
\includegraphics[width=0.85\linewidth]{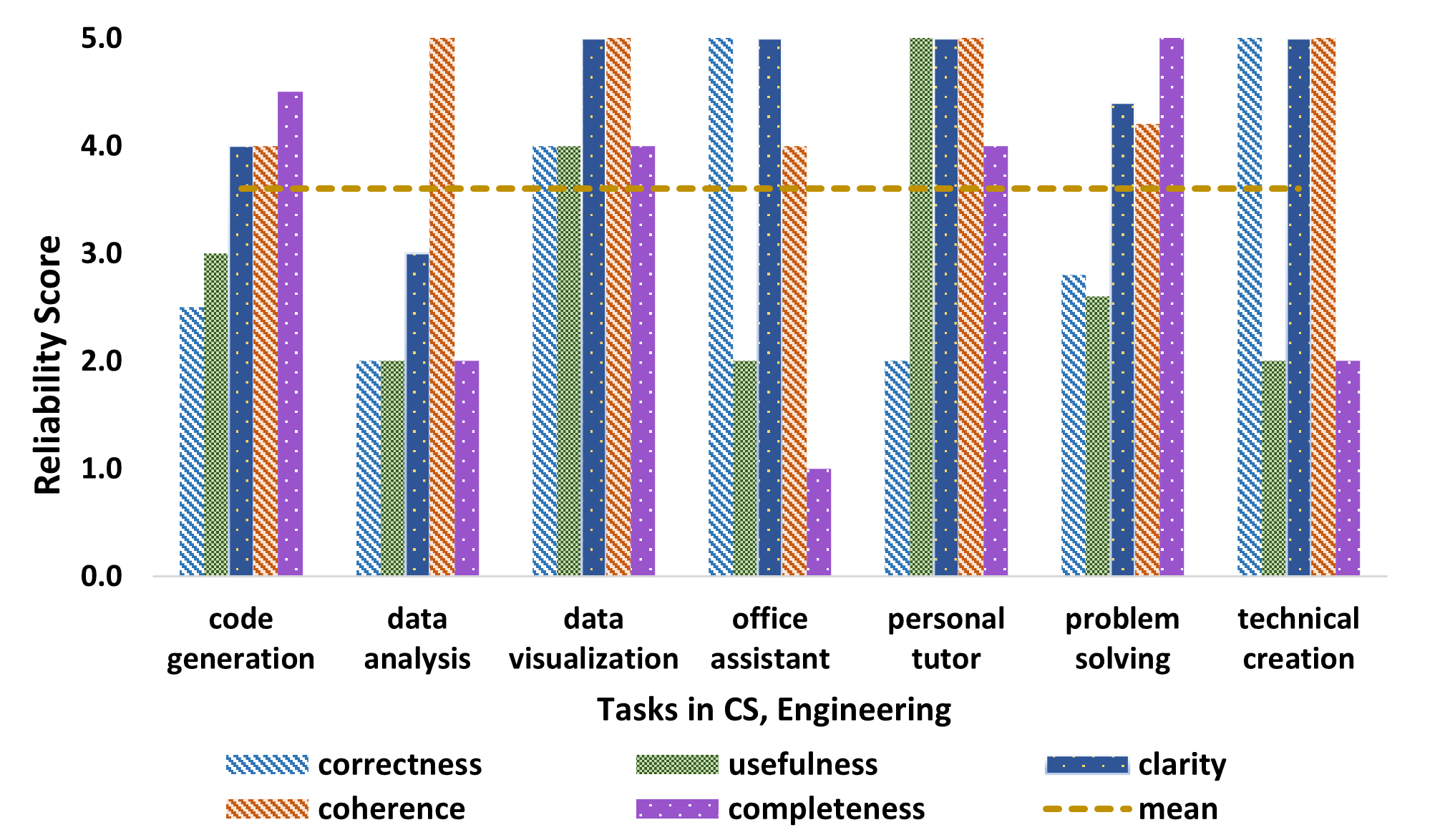}
\caption{Reliability score by five factors in CSE tasks; line chart shows the average reliability score per task type}
\label{fig:ana-by-task}
\end{figure}

\begin{figure}[!t]
\centering
\includegraphics[width=0.65\linewidth]{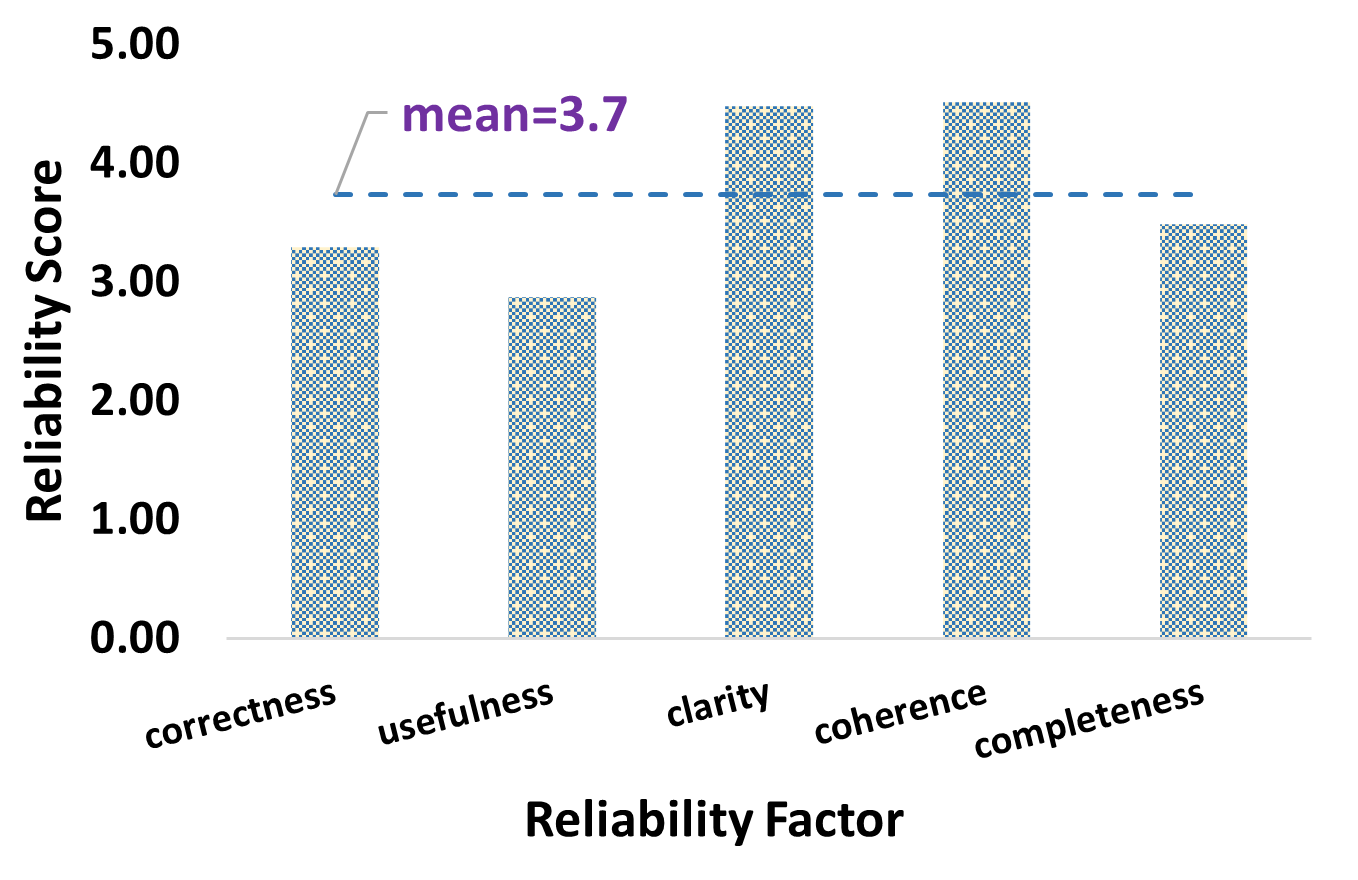}
\caption{Average score of five factors in CSE unsatisfied responses}
\label{fig:5-factor}
\vspace{-1ex}
\end{figure}

Figure \ref{fig:ana-by-subject} illustrates the reliability scores across five factors—correctness, usefulness, clarity, coherence, and completeness—within different subjects. Figure \ref{fig:ana-by-task} showcases how ChatGPT's reliability varies across task types, from coding tasks to providing personalized tutoring, focusing on these same five factors. In addition, Figure \ref{fig:5-factor} depicts the average score for each of the five reliable factors. Our primary observations  across these results include:

\begin{figure}[!t]
\centering
\includegraphics[width=0.95\linewidth]{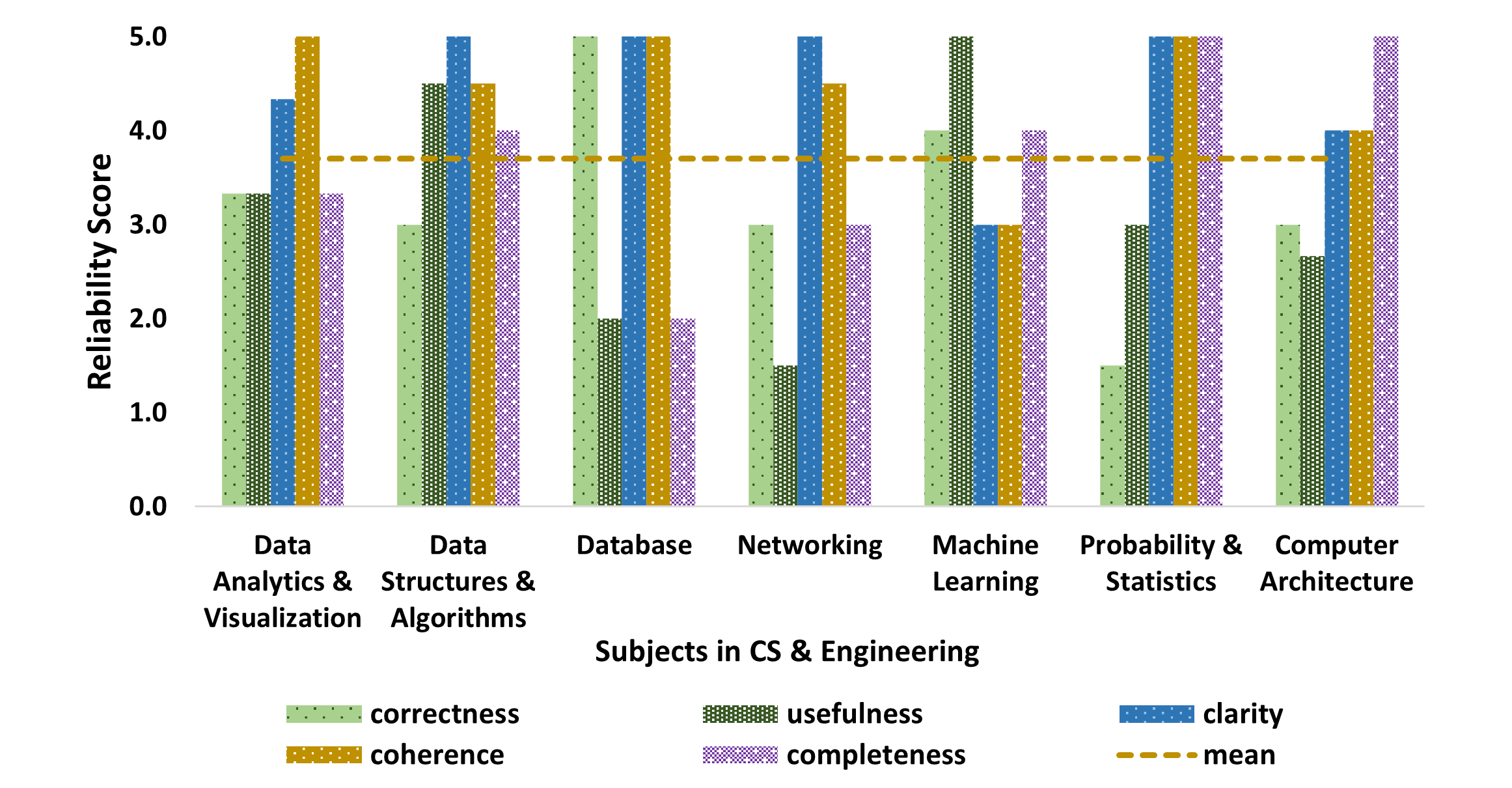}
\caption{Reliability score by five factors in CSE subjects; line chart shows the average reliability score per subject}
\label{fig:ana-by-subject}
\vspace{-1ex}
\end{figure}

\begin{itemize}[leftmargin=*]
    \item Among the five reliability factors, 
    usefulness and correctness pose the most significant challenges, followed by completeness, which is far below the mean score as shown in Figure \ref{fig:5-factor}. These are essential metrics for users' trust in ChatGPT. 
    \item Overall, ChatGPT performs well in clarity and coherence by subject and task type. 

\item Responses perceived as useful due to their clarity and coherence often fall short in correctness or completeness, impacting their overall reliability value. 
\item Correctness depends on task types. Code generation requires precise code, while data analysis needs content accuracy and suitable visual presentations. Diagram creation, like Entity Relationship Diagram (ERD), necessitates detailed explanations and proper inclusion of diagrams. 
    \item ChatGPT performs well as a personal tutor. Even though it fails to provide the correct answer in some tasks, users perceive its provided information as useful.
    \item It performs fairly well in generic data visualization but encounters occasional environment and error message issues. 
    \item Some codes generated by ChatGPT might hide errors, becoming apparent only when executed. While excelling in generating code for traditional subjects like data structures and algorithms, concealed errors might occur in ML tasks. 
    \item In technical design and creative tasks that require applying multiple subjects, ChatGPT's responses are often overly simplistic, and underperforming compared to search.
    \item ChatGPT performs well in table creation, while PowerPoint slide generation proves challenging for ChatGPT. 
\end{itemize}

\subsection {Exploring Unsatisfactory Responses: Case Studies}
\label{subsec:examples}
This section aims to analyze ChatGPT's weaknesses, identify causes, and suggest improvements for enhanced educational integration. 

\subsubsection{Hidden Errors in Code}

ChatGPT generally performs well in providing accurate answers in computer science and engineering topics, showcasing considerable knowledge and attention to detail. However, we observed instances where the code suggested by ChatGPT contained hidden errors that were challenging to identify upfront. In particular, in some cases, ChatGPT can suggest a chunk of code, that some lines of code are incorrect, or partially incorrect. 

\begin{figure}[!t]
\centering
\includegraphics[width=\linewidth]{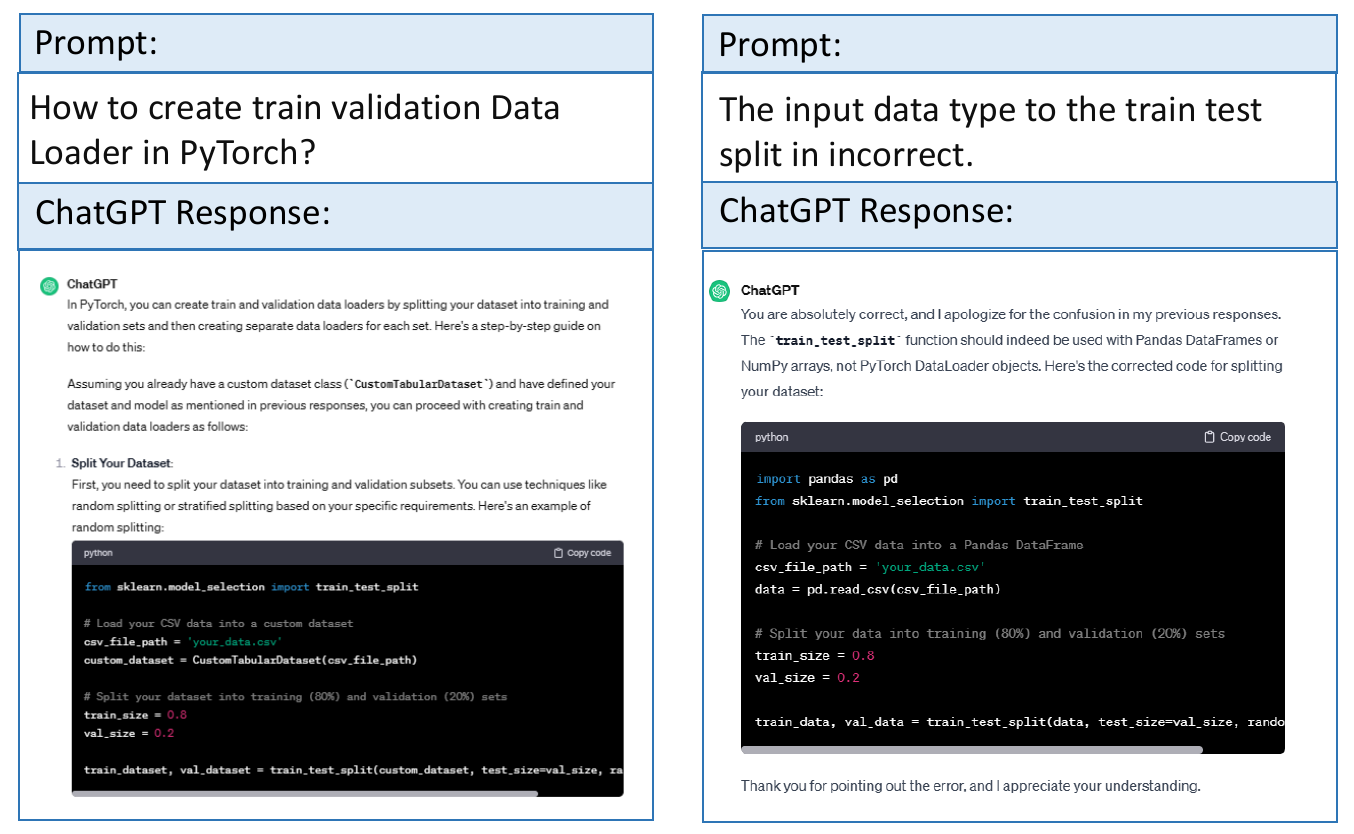}
\caption{Hidden errors (prompt \#7 in Table \ref{tab:reliability-score}).}
\label{fig:pytorch-dataloader}
\vspace{-1ex}
\end{figure}

For example, as shown in Figure \ref{fig:pytorch-dataloader}, initially, we engaged ChatGPT to aid in constructing a Convolutional Neural Network (CNN) model using PyTorch. 
After executing the supplied code on our dataset, we sought guidance on customizing the PyTorch dataset class for tabular data within a CSV file comprising 10,000 rows and 4 columns as features. ChatGPT provided code for a customized data class in PyTorch, seemingly accurate. However, upon attempting to run the code for creating a train-validation dataset, as depicted in the initial prompt in Figure \ref{fig:pytorch-dataloader}, errors surfaced. 

Upon inspection, we realized that the suggested approach by ChatGPT, while commonly used for train-validation splits, did not function correctly within the context of a customized dataset's Dataloader in PyTorch. Recognizing this discrepancy, we found that employing a different method (such as 'random\_split' or 'SubsetRandomSampler') from the PyTorch library was necessary for successful execution.
Our interaction with ChatGPT effectively guided us 
developing a CNN model. 
However, it is notable that despite its valuable assistance, hidden errors might exist highlighting the possibility of overlooking crucial details. Notably, ChatGPT omitted the necessity to encode categorical label data into a numerical format—a requisite for the dataloader to function correctly. In subsequent interactions, when we flagged the code's inaccuracies, ChatGPT acknowledged the issue and provided a corrected version of the code.

\textit{Assessment:} ChatGPT showcases a range of cognitive abilities, including memorization, understanding, and higher-level thinking to apply acquired knowledge. Notably, ChatGPT offers a comprehensive set of codes that function effectively, streamlining the process of searching and comprehending information. However, while the solutions provided by ChatGPT are generally applicable, they may encounter difficulties in adapting to specific use cases. Hence, users should exercise caution when implementing the suggested code.

\textit{Recommendations:} ChatGPT offers a quicker way to grasp new concepts, but educators must recognize its limitations. While it accelerates information gathering, it can't replace higher-level skills like critical thinking. Combining its data with traditional resources is beneficial, yet true learning demands more than just information absorption. Relying solely on ChatGPT for learning poses risks, as demonstrated in a case study.
Learners may use ChatGPT as a starting point for basic concepts. They should further deepen their understanding through critical analysis and creativity. By doing so, they turn ChatGPT's mistakes into opportunities for cognitive growth. However, spotting these errors might challenge undergraduates still mastering the fundamentals.
Meanwhile, it provides an opportunity for educators to use unsatisfied responses 
as lesson materials and engage students with collective inquiries with step-by-step guidance in the classroom. For ChatGPT improvement, continuous training and collecting users' feedback for better problem-solving in specific situations is crucial.

\subsubsection{Weak Visualization Skill}
\begin{figure}[!t]
\centering
\includegraphics[width=\linewidth]{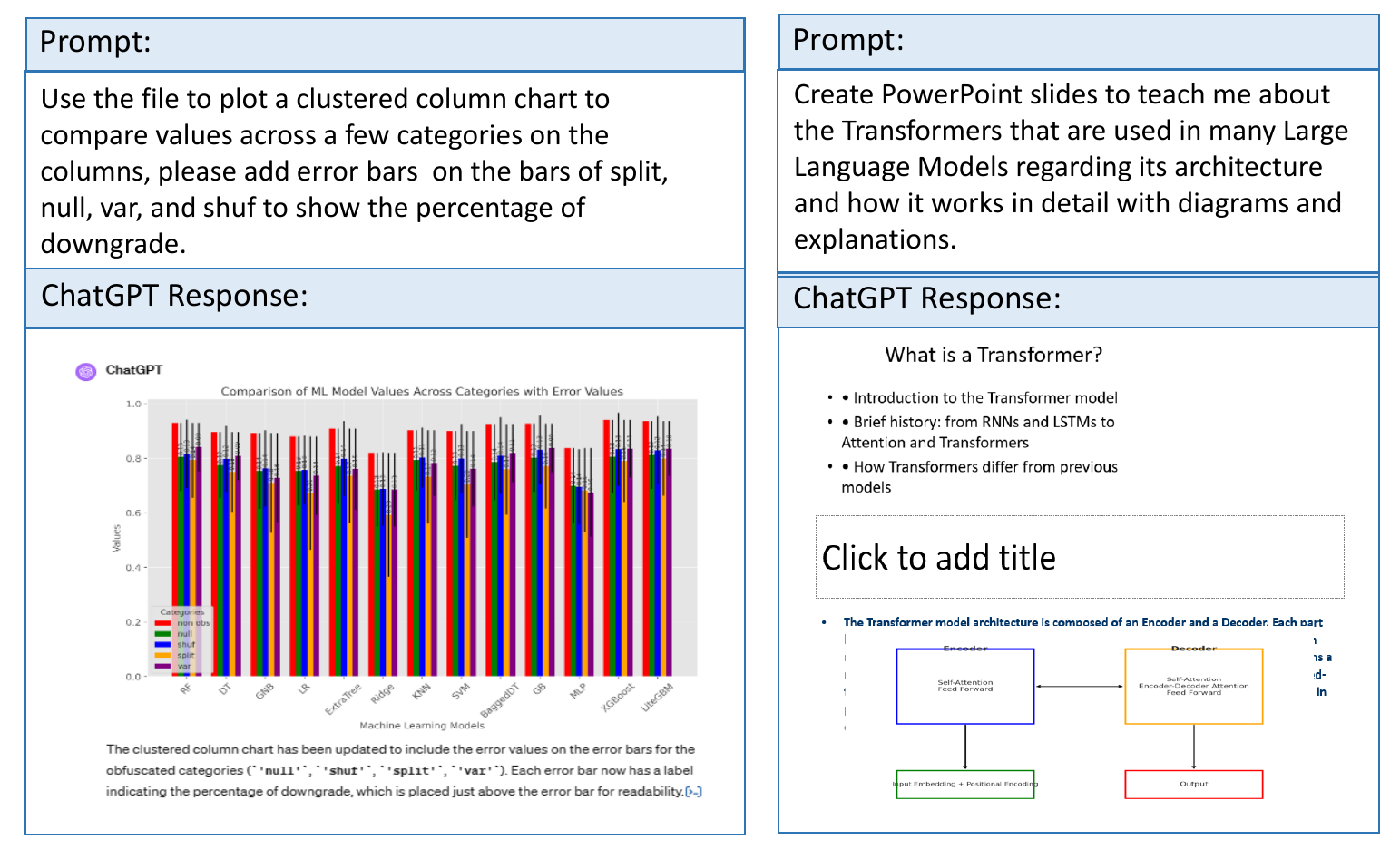}
\caption{Left: Plot a clustered column bar chart with error bars (prompt \#5, Table \ref{tab:reliability-score}); Right: Create a PPT with diagram describing Transformer architecture (prompt \#15, Table \ref{tab:reliability-score}).}
\label{fig:visualization-error}
 \vspace{-2ex}
\end{figure}

We tasked ChatGPT with different data visualization assignments. Initially, we requested clustered column charts with error bars, depicted in the first prompt of Figure \ref{fig:visualization-error}. While ChatGPT showed a good understanding of the chart type, the plotted graph had inaccuracies—the error arrows did not correspond correctly to the error values and were inconsistently placed for all bars.
In the second prompt, we asked ChatGPT to create PowerPoint slides explaining how Transformers work. Although it could generate extensive text content upon request for a description, its performance in creating slides was notably disappointing. 
The generated content lacked depth and completeness, resembling a preliminary outline drafted in a rush. 
Despite seeking further elaboration on Transformers' architecture, ChatGPT struggled to provide useful information. When prompted to include an architecture diagram, it produced a very abstract representation, affected by display issues such as overlapping elements and improper positioning of figures and text in the PowerPoint. 
From our observation, ChatGPT seems to generate diagrams based on its recall of abstract-level processes, attempting to code the components without delving into the 
detailed workflow of the architecture.

\textit{Assessment: }
ChatGPT shows entry-level skills in data visualization. It struggles compared to humans using tools like Excel or PPT, where the human ability to combine existing and new knowledge surpasses its limitations. ChatGPT lacks seamless integration of its knowledge with external sources like search engines. Yet, it excels when prompted to generate code for standard graphs using libraries like scikit-learn and Matplotlib. Human creators can outperform ChatGPT in more complex tasks like multi-variant analysis and graph generation.

\textit{Recommendations:} Data visualization can be challenging due to its artistic nature and the need to convey meaningful information on the graph.  ChatGPT can help by offering various visualization options and explanations. However, its limitations create an opportunity for educators. They can use ChatGPT's suggested plots as case studies in data analysis courses, discussing strengths and weaknesses with students to enhance their skills in this field. Some suggestions for enhancement include incorporating an option for ChatGPT to save the generated data and chart into an Excel file, allowing further editing. Additionally, enabling a search function could assist in consolidating information from various sources, enhancing the quality of generated slides. Training ChatGPT to understand blogs and concepts, could provide more insightful information. To address potential plagiarism concerns, ChatGPT could be trained to cite sources used in its responses.

\subsubsection{Oversimplified and Ineffective Response}
\begin{figure}[!t]
\centering
\includegraphics[width=\linewidth]{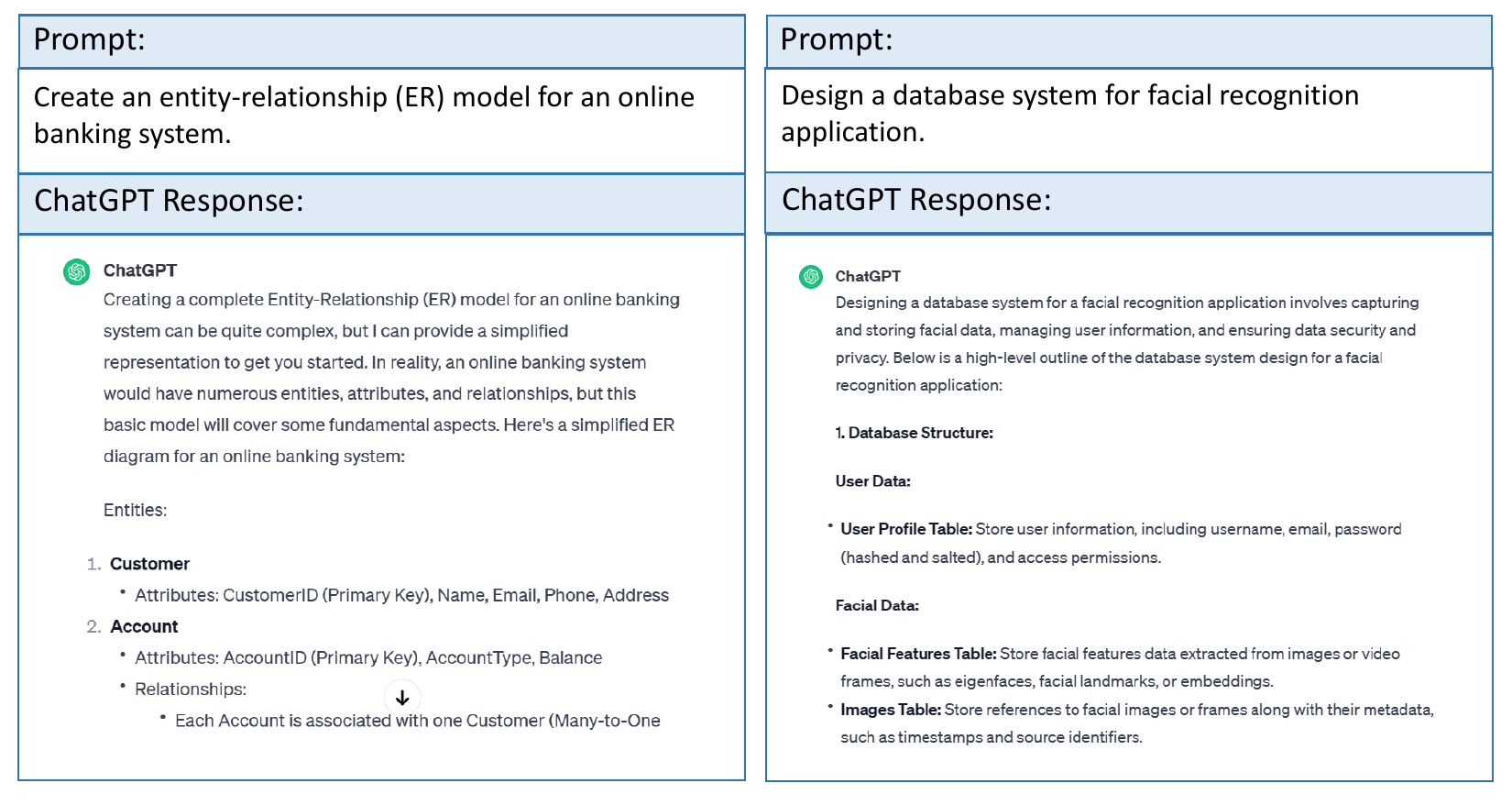}
\caption{Left: Create ERD model for online-banking system (prompt \#8, Table \ref{tab:reliability-score}); Right: Design DB for facial recognition (prompt \#9, Table \ref{tab:reliability-score})}
\label{fig:too-simple}
\end{figure}

We found that some of ChatGPT's responses were overly simplistic, and lacking practical usefulness and utility. For instance, as shown in Figure \ref{fig:too-simple} when prompted to create an Entity-Relationship (ER) model for an online banking system, as commonly encountered in CS courses, ChatGPT generated a simple yet technically correct response. For this question, providing an ER diagram is essential. However, the lack of an expected ER diagram diminished its usefulness. In contrast to online resources that detail such designs with labeled cardinalities and relationship types, ChatGPT's performance fell relatively short of user expectations. 
Our second inquiry involves asking ChatGPT to design a database system for facial recognition applications. As depicted in the second prompt in  Figure \ref{fig:too-simple}, the response merely presents a broad overview of the database system design without offering detailed guidance, code samples, or thorough explanations. Comparable online educational materials are notably more informative. Despite the abundance of available public content on the internet, it appears that ChatGPT cannot synthesize its existing knowledge into a comprehensive system design. It currently operates without a holistic intelligence, accumulating fragmented knowledge.

\textit{Assessment: }These questions demand high-order cognitive skills that involve applying, synthesizing, and creating knowledge. ChatGPT's performance in this evaluation was lacking.

\textit{Recommendations:}  ChatGPT could benefit from integrating search engine results for such queries. While improving its training in these tasks might enhance proficiency, developing higher cognitive skills could take time. For users encountering gaps in ChatGPT's knowledge, we recommend turning to stable sources like search engines, as they often offer more comprehensive information than ChatGPT in specific cases.

\subsubsection{Inconsistency between Memorization and Reasoning}
\begin{figure}[!t]
\centering
\includegraphics[width=\linewidth]{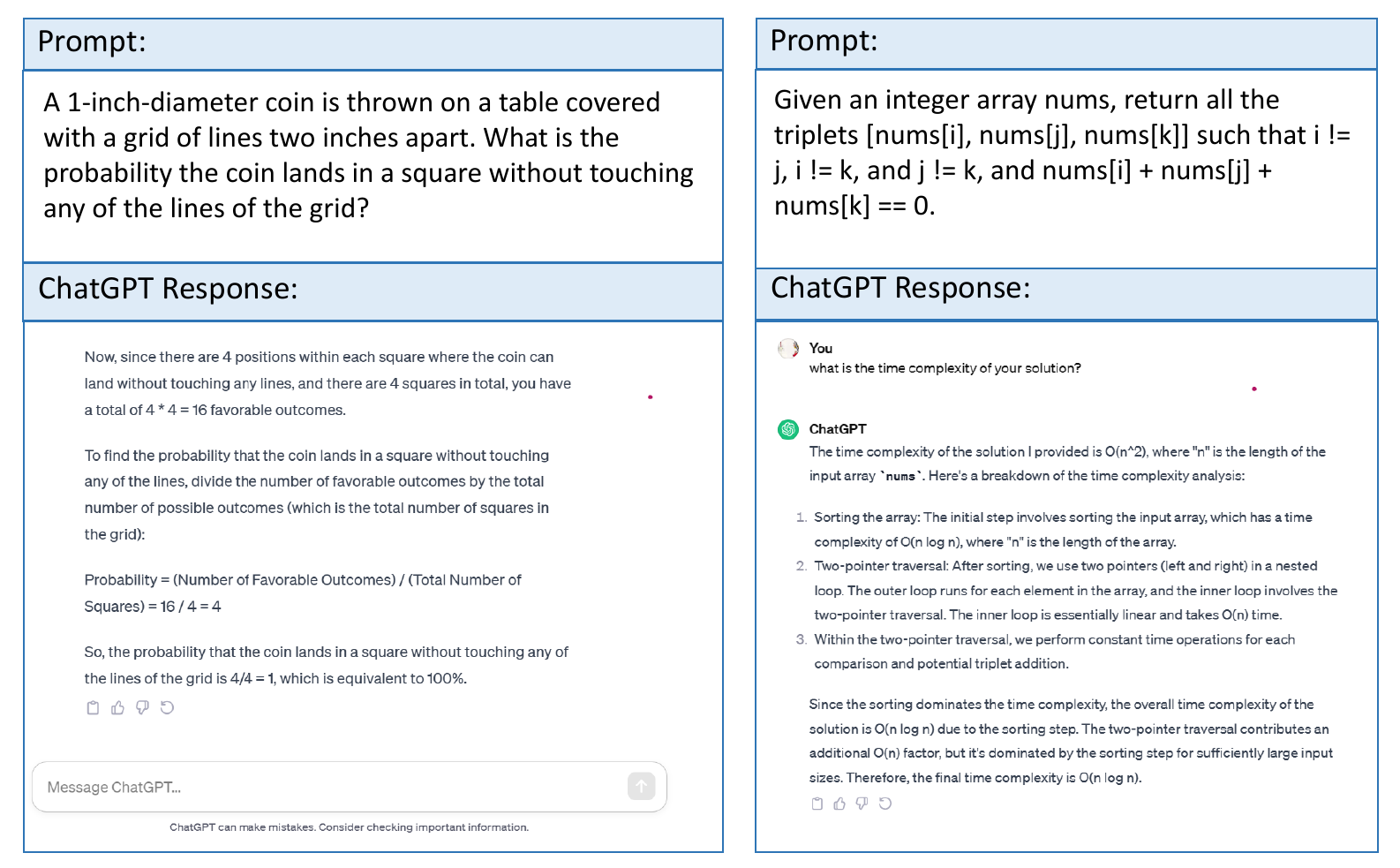}
\caption{Left: Coin probability (prompt \#10 in Table \ref{tab:reliability-score}); Right: Time complexity analysis (prompt \#11 in Table \ref{tab:reliability-score})}
\label{fig:inconsistency}
\vspace{-1ex}
\end{figure}

ChatGPT exhibits inconsistency between memorization and logical reasoning. In the scenario depicted in Figure \ref{fig:inconsistency}, we initiated ChatGPT with a probability question. The initial response (left screenshot) presented a lengthy logical reasoning that initially seemed accurate but concluded with an incorrect answer.
Subsequently, we presented an algorithm problem-solving query (right screenshot). ChatGPT accurately resolved the algorithmic challenge by providing the correct code. Following this, we inquired about its understanding of time complexity. While initially providing a correct answer, ChatGPT subsequently drifted during the reasoning process, resulting in an incorrect and disparate conclusion.

\textit{Assessment: }ChatGPT demonstrates the utilization of different sets of cognitive skills when addressing a single problem. This disparity results in conflicting answers, highlighting a difference from human consistency in synthesizing diverse skill sets. ChatGPT operates similarly to a collective thinking process, where different  "teams" may yield different answers.

\textit{Recommendations: }Training ChatGPT to apply higher-order cognitive skills presents a significant challenge. Implementing a sanity-checking mechanism within ChatGPT's response generation process could ensure consistency, synthesizing diverse outputs into a unified and coherent response. This case study highlights the risks of relying solely on ChatGPT's answers in CSE and beyond. While it's a valuable learning tool, its reasoning process needs to be validated.  
Educators can use ChatGPT's flaws as case studies, encouraging students to critically evaluate responses. Shifting from dominating teaching concepts to exploring areas for improvement fosters critical thinking and higher cognitive skills in students.

\subsubsection{Naive Answers in Computer Architecture \& Organization}

\begin{figure}[!t]
\centering
\includegraphics[width=0.99\linewidth]{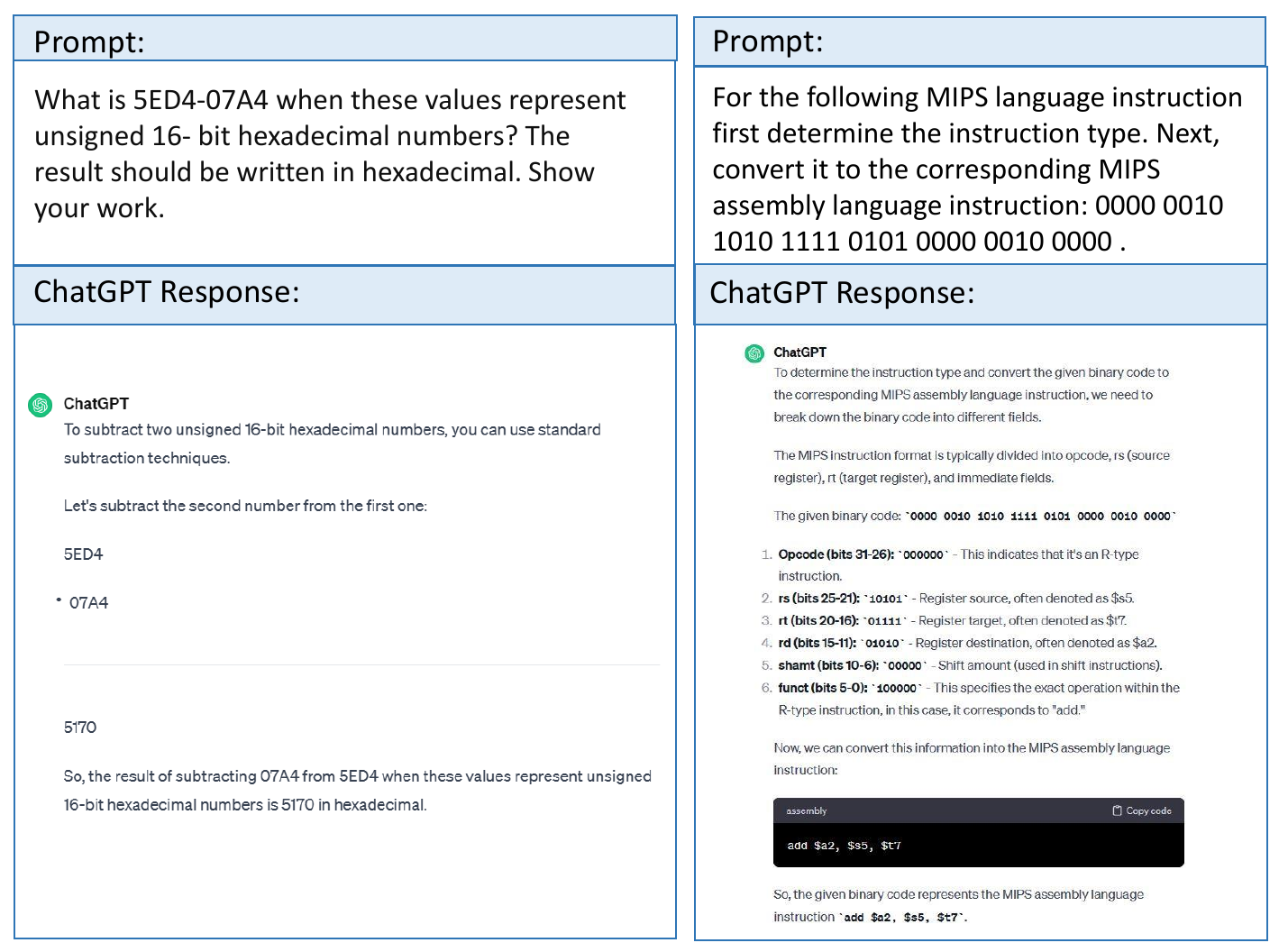}
\caption{Left: binary arithmetic (prompt \#12 in Table \ref{tab:reliability-score}); Right: MIPS instruction (prompt \#14 in Table \ref{tab:reliability-score})}
\label{fig:binary-substraction}
\vspace{-1ex}
\end{figure}

ChatGPT often faces limitations in foundational binary arithmetic operations and MIPS instruction conversion queries in computer architecture subjects. In Figure \ref{fig:binary-substraction}, when prompted with a basic binary operation and an MIPS binary-to-assembly conversion, ChatGPT provided two incorrect answers.

\textit{Assessment:} ChatGPT accurately identified most MIPS assembly language instructions but mistakenly mixed up registers \$t2 and \$a2. Notably, ChatGPT showcased its capacity for self-correction upon the error being pointed out, indicating its cohesive ability but needed multiple attempts in a new chat to achieve this. In another case, field segmentation errors led to wrong register use or instruction identification. ChatGPT's responses vary and might be impacted by past interactions, sometimes resulting in inaccuracies or confusion.  
For reliable and consistent results, starting a new session is recommended.

\textit{Recommendations: } Exercise caution when using ChatGPT for logic design and computer architecture queries, as errors may occur. Consider ChatGPT as a reference tool rather than the sole solution. Educators can leverage incorrect responses for critique, group discussions, and error correction exercises. Supplementing ChatGPT with hands-on practice using reliable educational resources and expert guidance enhances learning reliability in these domains.

\subsubsection{Technical Glitches}

\begin{figure}[!t]
\centering
\includegraphics[width=.7\linewidth]{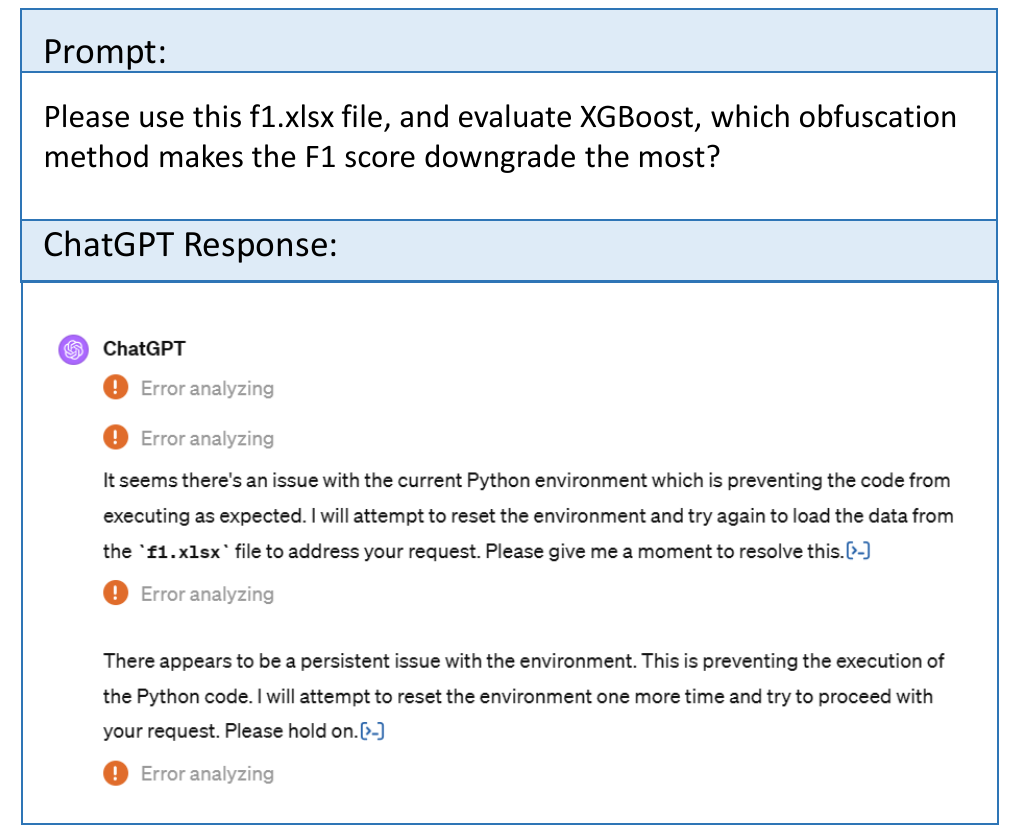}
\caption{Environment issue (prompt \#6 in Table \ref{tab:reliability-score}).}
\label{fig:technical-glitch}
\vspace{-1ex}
\end{figure}

We occasionally found that ChatGPT encounters technical glitches, such as environment issues, resource constraints in the environment, code execution environment issues, and connection issues.  In a case shown in Figure \ref{fig:technical-glitch}, we tasked ChatGPT with analyzing a file for code obfuscation methods, where it repeatedly faced Python environment challenges. These technical glitches were encountered occasionally across our experiments with ChatGPT.

\textit{Assessment: } Overall, ChatGPT operates within a stable environment for task execution. However, occasional technical glitches arise, particularly evident in visualization inquiries.  

\textit{Recommendations: } Enhancements in server capacity, ensuring robust connectivity, and optimizing the execution environment might help alleviate these challenges. Additionally, regular maintenance, updates, and ensuring compatibility with various file formats and environments could contribute to minimizing these technical glitches and stabilizing the code environment. Temporary technical glitches can often be resolved by regenerating results, starting a new conversation thread, or restarting ChatGPT. However, persistent issues might require seeking information from alternative sources after using ChatGPT's response as an initial reference.

\subsection{Evaluation of ChatGPT's Cognitive Skills}
\label{sec:Bloom}
\begin{figure}[!t]
\centering
\includegraphics[width=0.95\linewidth]{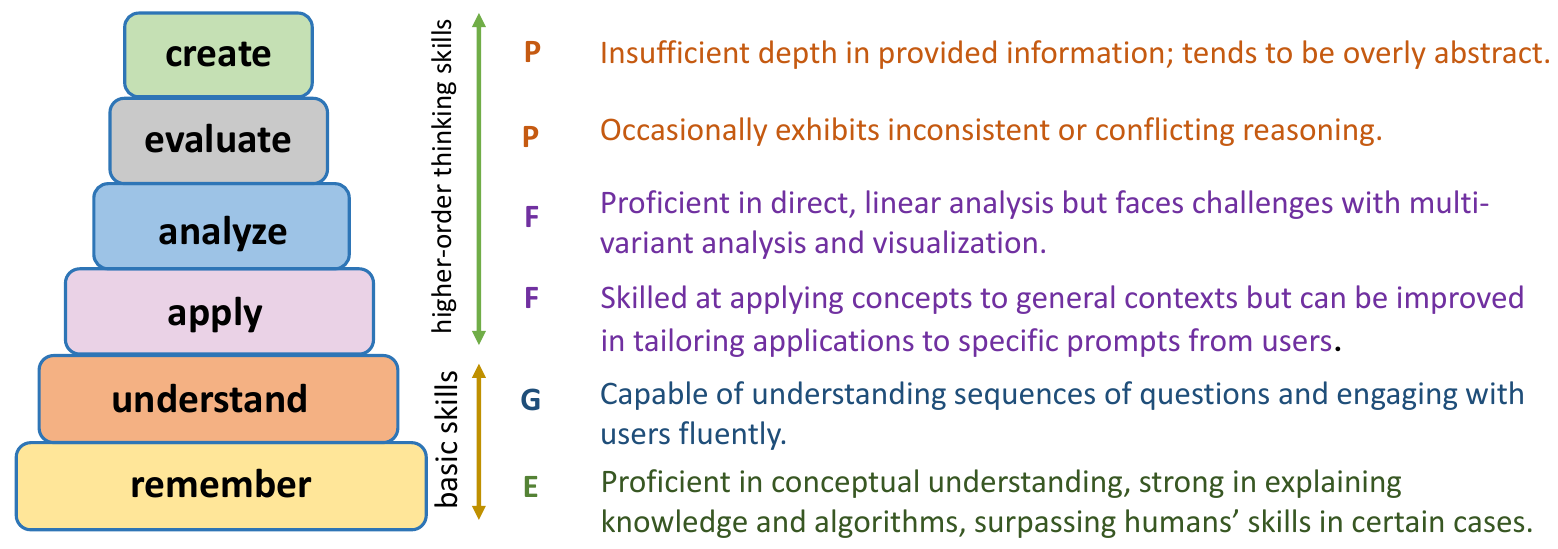}
\caption{Evaluating ChatGPT's competence of cognitive skills according to Bloom’s taxonomy (E: Excellent, G: Good, F: Fair, P: Poor)
}
\label{fig:cognitive}
\vspace{-2ex}
\end{figure}

We assess ChatGPT's cognitive capabilities through the lens of Bloom's taxonomy \cite{anderson2001taxonomy} to gauge its learning levels and complexity of reasoning. Evaluating these skill sets assists educators and learners in understanding ChatGPT's 
strengths and weaknesses. Employing a scale of Excellent (E), Good (G), Fair (F), and Poor (P), our assessment, illustrated in Figure \ref{fig:cognitive}, reflects ChatGPT's proficiency across various cognitive skills. 
ChatGPT excels in basic skills—ranking Excellent in Remember and Good in Understand. It excels in applying and analyzing complex concepts but faces limitations in translating knowledge to unique problem-solving contexts, notably in areas such as data analysis, visualization, and computer architecture.  
For Evaluate and Create, ChatGPT exhibits poor performance, demonstrating shallow thinking, lack of depth, and inconsistency in reasoning, leading to conflicts in responses; thus, these skills rank as Poor.

\section{Implications in Learning and Assessment}
\label{sec:implication}

This section outlines ChatGPT's impact in education, addressing reliability, fairness, and integrity. 
We further suggest potential adaptations of higher education in the ChatGPT era.

\vspace{0.3ex}
\noindent
\textit{- Reliability:} Our experiments demonstrated ChatGPT's inherent unreliability in some subjects, manifesting in errors, incomplete or inconsistent reasoning, and misleading content presented with a polished structure. This unreliability undermines students' learning outcomes and raises open and complex questions. 
Differentiating acceptable content from seemingly flawless responses necessitates solid subject knowledge. However, for undergraduates studying computer science and engineering and learning foundational concepts, recognizing these flaws presents a significant challenge. Considering ChatGPT as an educational tool within the course curriculum, the responsibility for detecting such issues remains unclear, potentially leading to confusion  
in educational settings.

\vspace{0.3ex}
\noindent
 \textit{- Fairness:} The biases and hidden errors in ChatGPT create unfair advantages for learners, impacting the overall learning outcomes. 
 Without clear policies on ChatGPT usage as an educational tool and assessment method, grading fairness becomes uncertain. Basic assignments where ChatGPT excels might lead to students overly relying on it, affecting their learning incentives.
 Also, unequal access to AI tools creates disparities among students, complicating assignment development and assessment fairness \cite{baird-FIE2023-race,fernandes2023advancing}.  
 This challenge demands curriculum adjustments and policy reform in higher education.

\vspace{0.3ex}
\noindent
\textit{- Integrity:} The emergence of ChatGPT raises serious concerns about academic integrity. Plagiarism becomes a pressing issue with using ChatGPT in education, risking the integrity of educational outcomes and degrees, and comprising the next generations’ cognitive skills. This highlights the need for higher education to develop policies that integrate advanced AI-based tools while protecting academic integrity. 
Moreover, without proper regulations, educators might solely rely on AI-generated content for teaching, neglecting higher-order skill development in students. Establishing new standards that account for the use of AI-based tools in education is imperative. 
We further consider the following implications and suggestions in response to the highlighted impacts on CSE education.

\vspace{0.3ex}
\noindent
\textit{- Adapting Policy:} As higher education increasingly embraces human-AI collaboration, it is vital for CSE programs to reassess policies regarding generative AI tools like ChatGPT. 
This involves redefining learning outcomes, ethical guidelines, and assessment methods, alongside providing campus-wide policy directives for colleges and departments. Key questions must be addressed, such as redefining plagiarism  
and determining the ethical boundaries \cite{DWIVEDI2023102642}. 
More funding and policy adjustments are 
essential to enable schools to explore, test, and adopt advanced AI tools,  enriching the learning experience. 

\vspace{0.3ex}
\noindent
    \textit{- Curriculum Shift:} Educational strategies and curriculum must shift from foundational skills to advanced cognitive abilities like application, analysis, and creation when integrating AI tools. 
    This transformation requires significant resources for training and experimentation, challenging educators to reconsider teaching methods by emphasizing critical thinking and practical application of knowledge. 
    For instance, in programming courses, students can be allowed to use AI-generated code within specific applications, fostering deeper understanding and problem-solving skills. Similarly, integrating real-world challenges into learning experiences enhances scientific understanding and soft skills like teamwork and communication \cite{CBL-2023}. The CSE programs' adaptability makes it ideal for multidisciplinary approaches, incorporating diverse subjects within courses to cultivate higher-order skills, with a focus on practical real-world case studies and applications. 

\vspace{0.3ex}
\noindent
\textit{- Revisiting Assessment:} It is crucial to redefine assessment strategies in the ChatGPT era, focusing on evaluating students' application of CS and engineering concepts rather than mere memorization. This entails incorporating assessments that cultivate active learning and encourage practical use of knowledge, leveraging AI-based tools for initial learning stages \cite{qureshi2023exploring}. Educators should adopt a mix of formative and summative assessments, employing personalized methods like presentations, hands-on labs, in-class activities, and critical thinking evaluations to thoroughly measure students' learning.

\section{Concluding Remarks}
\label{sec:con}

In this study, we delved into ChatGPT's potential in Computer Science and Engineering (CSE) education, exploring its strengths and limitations. 
Employing a systematic approach, we create diverse challenging educational problems focusing on personalized, in-depth questions and project tasks to assess the reliability of ChatGPT's responses.  
We introduced a reliability analysis framework to evaluate the ChatGPT's performance across five deterministic factors.  
Our findings highlight machine learning, probability and statistics, and computer architecture and organization as the most challenging subjects for ChatGPT, while data analysis proves to be its toughest task. Despite impressive code generation capabilities, unique problem-solving remains a challenge. 
Our comprehensive analysis sheds light on both the effectiveness and challenges of ChatGPT within CSE education, underscoring the need for continuous refinement across the tool, curriculum, and institutional policies to facilitate effective and in-depth learning experiences. This offers crucial guidance for educators and learners, aiding in navigating its reliability factors to prevent potential adverse impacts on education for future generations.

\begin{small}

\bibliography{literature} 
\end{small}

\end{document}